\documentclass[sn-mathphys-num]{sn-jnl}

\usepackage{graphicx}%
\usepackage{multirow}%
\usepackage{amsmath,amssymb,amsfonts}%
\usepackage{amsthm}%
\usepackage{mathrsfs}%
\usepackage[title]{appendix}%
\usepackage{xcolor}%
\usepackage{textcomp}%
\usepackage{manyfoot}%
\usepackage{booktabs}%
\usepackage{algorithm}%
\usepackage{algorithmicx}%
\usepackage{algpseudocode}%
\usepackage{listings}%
\usepackage{pdflscape}
\usepackage{graphicx}
\usepackage{placeins}
\def\bm#1{{\mbox{\boldmath{$#1$}}}}

\theoremstyle{thmstyleone}%
\theoremstyle{thmstyletwo}%

\theoremstyle{thmstylethree}%

\begin{document}

\title[Variable Selection in Two-stage Joint Modeling]{\textbf{A Two-Stage Bayesian Approach for Variable Selection in Joint Modeling of Multiple Longitudinal Markers with Competing Risks}}


\author*[1,2]{Taban Baghfalaki}\email{taban.baghfalaki@manchester.ac.uk}
\author[3]{Reza Hashemi}
\author[2]{Christophe Tzourio}
\author[2]{Catherine Helmer}
\author[2]{Helene Jacqmin-Gadda}
\affil[1]{\orgdiv{Department of Mathematics}, \orgname{The University of Manchester}, \orgaddress{\street{Manchester},  \country{UK}}}
\affil[2]{\orgdiv{Inserm, Research Center U1219, Univ. Bordeaux}, \orgname{ISPED}, \orgaddress{\street{F33076 Bordeaux},  \country{France}}}
\affil[3]{\orgdiv{Department of Statistics}, \orgname{Razi University}, \orgaddress{\street{Kermanshah}, \country{Iran}}}

\abstract{
In many clinical and epidemiological studies, collecting both longitudinal measurements and time-to-event outcomes is essential. Accurately estimating the relationships between longitudinal markers and time-to-event outcomes, as well as identifying key longitudinal markers for risk prediction, is critical, especially in the presence of {competing risks}, where multiple types of events can occur. However, as the number of longitudinal markers increases, fitting joint models becomes increasingly challenging, often resulting in long computation times or convergence issues.  
In this paper, we propose a novel two-stage Bayesian approach for variable selection in joint models that accommodate multiple longitudinal markers and {competing risks outcomes}. Our method is designed to efficiently identify significant longitudinal markers and covariates. In the first stage, we fit a one-marker joint model for each longitudinal marker together with the {competing risks event}. These estimates are then used to predict individual marker trajectories, mitigating bias caused by informative dropout. In the second stage, we fit a cause-specific hazards model for competing risks, incorporating the predicted current values of all markers as time-dependent covariates.
We explore both continuous and Dirac spike-and-slab priors for variable selection within a Bayesian framework, using Markov chain Monte Carlo (MCMC) techniques. Our approach enables risk prediction based on a large number of longitudinal markers—a task that is often considered infeasible in standard joint modeling approaches. We rigorously evaluate our method through simulation studies, assessing both parameter estimation and predictive performance.  
Finally, we illustrate the practical utility of our approach by predicting the risk of dementia using the Three-City (3C) dataset, a longitudinal cohort study conducted in France, where {competing risks of death are present}. To facilitate implementation, we provide an R package, \texttt{VSJM}, freely available at \url{https://github.com/tbaghfalaki/VSJM}.  
}


\maketitle

\section{Introduction}\label{sec1}
In many clinical studies, longitudinal measurements are repeatedly collected for each subject at different time points until an event such as death or drop-out occurs. It is also common to have multiple longitudinal measurements in a study. For example, in the context of HIV/AIDS studies, two important markers that are often repeatedly measured until death or drop-out are HIV-RNA and CD4 \citep{guedj2011joint}. In most of these types of studies, the number of longitudinal markers under investigation and the number of explanatory variables is large. Estimating the association between longitudinal markers and time-to-event outcomes, as well as determining the effect of explanatory variables on event risks, is crucial in modern medicine \citep{hossain2020multivariate,shen2021backward,lin2022deep,devaux2022random,zhang2023multivariate}, in particular, for the development of individual risk prediction models  \citep{rizopoulos2011dynamic,rizopoulos2012joint}.
For this purpose, joint models for longitudinal and time-to-event outcomes
that combine mixed models and time-to-event models,
have received considerable attention over the past decades. 
Interesting reviews on joint modeling can be found in \cite{sousa2011review,papageorgiou2019overview,alsefri2020bayesian} and \cite{zhudenkov2022workflow}. 
From a computational point of view, the R software now offers some options: \texttt{JMBayes} \citep{rizopoulos2014r}, \texttt{JMBayes2} \citep{rizopoulos2022jmbayes2},  \texttt{INLAjoint} \citep{rustand2022denisrustand}, and stan\_jm modelling function in the \texttt{rstanarm} package \citep{brillemanestimating}. The Bayesian implementation of joint models with multiple markers is available through \texttt{INLAjoint} \citep{rustand2022denisrustand} and \texttt{JMBayes2} \citep{rizopoulos2022jmbayes2}, but dealing with a large number of markers still poses computational difficulties due to the large number of random effects and parameters in these models.
For this purpose, \cite{Baghfalakitsjm} propose a novel two-stage joint modeling approach (TSJM). In the first stage, separate joint models are fitted for each biomarker and the time-to-event outcome. In the second stage, the predicted linear predictors from each biomarker-specific model are incorporated as time-dependent covariates in a semi-parametric Cox model. This strategy improves computational tractability and mitigates bias arising from informative dropout.
In their study, \cite{Baghfalakitsjm} compared the TSJM with the standard two-stage (STS) approach and a multi-marker joint model (MMJM), evaluating performance in terms of parameter estimation and dynamic prediction. Their results indicate that when the number of biomarkers is small, TSJM performs comparably to MMJM, while the STS approach performs poorly, particularly when the biomarkers are strongly correlated. Furthermore, when the number of biomarkers is large, MMJM exhibits limited estimation reliability, whereas TSJM provides a more favorable balance between estimation accuracy and model stability.
\\
When   multiple longitudinal markers and large number of time-invariant explanatory variables, are considered, the selection of the most predictive variables is challenging. Traditionally this has been performed through model comparison criteria such as Akaike Information Criterion (AIC), Bayesian Information Criterion (BIC), or Deviance Information Criterion (DIC). The use of these criteria for complex models in the presence of a large number of candidate models is not feasible due to the computational complexity associated with evaluating these criteria for complex models with numerous parameters and candidate models. Consequently, traditional criteria may not effectively guide model selection in such complex scenarios. This has led to the exploration of alternative strategies, such as variable selection, to address these challenges and identify the most appropriate model.
 It involves choosing appropriate variables from a complete list and removing those that are irrelevant or redundant. The purpose of such selection is to determine a set of variables that will provide the best fit for the model in order to make accurate predictions \citep{chowdhury2020variable}. In general, variable selection in the frequentist paradigm involves the utilization of a penalized likelihood approach, which considers various forms of penalty functions. These include the least absolute shrinkage and selection operator (LASSO), the smoothly clipped absolute deviation (SCAD), and the adaptive least absolute shrinkage and selection operator (ALASSO)  \citep{buhlmann2011statistics}. 
In the Bayesian framework, the selection of appropriate priors plays a pivotal role, similar to the regularization and shrinking techniques employed in frequentist variable selection methods. Properly chosen priors not only inform the prior beliefs about model parameters but also contribute to the regularization of the model, aiding in the prevention of overfitting and the promotion of model parsimony \citep{gelman1995bayesian}.
In this paper, we delve into the realm of Bayesian variable selection, where the careful selection of priors allows for the automatic identification of relevant predictors while appropriately handling model complexity \citep{george1993variable}. Various types of priors have been proposed for variable selection, including spike-and-slab priors \citep{malsiner2018comparing,george1993variable,mitchell1988bayesian}, horseshoe prior \citep{carvalho2010horseshoe}, and Laplace (i.e., double-exponential) prior \citep{williams1995bayesian}.
Our investigation focuses on two prominent types of priors: continuous and Dirac spike priors. These priors strike a balance between flexibility and simplicity, offering practical advantages in the context of variable selection \citep{scott2010bayes}. By exploring the implications and effectiveness of these priors, we aim to provide insights into their utility and contribute to the ongoing discourse surrounding Bayesian variable selection methodologies.\\
Variable selection methods in joint modeling have been explored across various studies, each offering unique insights into different aspects of the modeling process. \cite{he2015simultaneous} employed a penalized likelihood approach with adaptive  LASSO penalty functions to select fixed and random effects in joint modeling of one longitudinal marker and time-to-event data. \cite{yi2022simultaneous} extended this approach to the joint analysis of longitudinal data and interval-censored failure time data, utilizing penalized likelihood-based procedures to simultaneously select variables and estimate covariate effects. \cite{griesbach2023variable} adopted gradient boosting for variable selection in joint modeling, providing an alternative method for addressing this task.
\cite{chen2017variable} utilized L1 penalty functions to select random effects and off-diagonal elements in the covariance matrix in the survival sub-model. Their focus was on joint modeling of multiple longitudinal markers and time-to-event data within the frequentist paradigm. \cite{tang2023variable} considered simultaneous parameter estimation and variable selection in joint modeling of longitudinal and survival data using various penalized functions.  \cite{xie2020variable} utilized a B-spline decomposition and penalized likelihood with adaptive group LASSO to select the relevant independent variables and distinguish between the time-varying and time-invariant effects for the two sub-models.
On the Bayesian side, \cite{tang2017bayesian} proposed a semiparametric joint model for analyzing multivariate longitudinal and survival data. They introduced a Bayesian Lasso method with independent Laplace priors for parameter estimation and variable selection. However, these studies primarily focused on scenarios with a limited number of longitudinal markers and did not address cases where the number of markers is high.
Additionally, dynamic prediction in the context of variable selection in joint modeling remains unexplored in the literature reviewed here.\\
In this paper, we consider variable selection, including the selection of  both the longitudinal markers and the time-invariant covariates, in the joint modeling of multiple longitudinal measurements and competing risks outcomes. This is the first study to specifically address variable selection in this context.
For this purpose, we are faced with two issues. The first challenge is estimating joint models including a large number of longitudinal markers, which may be unfeasible. The second challenge, which involves variable selection, is addressed by using a spike-and-slab prior in a Bayesian paradigm.
As a summary, we propose a new two-stage approach for variable selection of multiple longitudinal markers and competing risks outcomes when the number of longitudinal markers and covariates is large, and the estimation of the full joint model including all the markers and covariates is intractable. Using a Bayesian approach, the first stage involves estimating one-marker joint models for the competing risks and each longitudinal marker as in \cite{Baghfalakitsjm}. This avoids the bias caused by informative dropout. In the second stage, extending TSJM, we estimate a cause-specific hazard model for the competing risks outcome using a Bayesian approach allowing variable selection using a Bayesian approach allowing variable selection. This model includes time-dependent covariates and incorporates spike-and-slab priors. The explanatory variables in the hazard model consist of the estimates of functions derived from the random effects and parameters obtained from the initial stage of jointly modeling each longitudinal marker.
We provide an R-package \texttt{VSJM} that implements the proposed methods which is available at \url{https://github.com/tbaghfalaki/VSJM}. \\
This paper is organized as follows: In Section 2, we describe our variable selection approach, which includes introducing longitudinal and competing risks sub-models, reviewing parameter estimation in MMJM, and finally  describing our proposed two-stage approach  for Bayesian variable selection. Section 3 focuses on dynamic prediction using the proposed approach. Section 4 presents a simulation study comparing the behavior of the variable selection two-stage approach with the oracle joint model in terms of variable selection and dynamic prediction.
In Section 5, the proposed approach is applied to variable selection and the risk prediction of dementia in the Three-City (3C) cohort study \citep{antoniak2003vascular}. Finally, the last section of the paper includes a conclusion of the proposed approach.
\section{Variable selection for joint modeling }
The observed time for subject \(i\) is defined as 
$T_i = \min(T_{i1}^*, \dots, T_{iL}^*, C_i)$,
where \(T_{il}^*\) denotes the time to event \(l\) and \(C_i\) is the censoring time. The corresponding event indicator is denoted by \(\delta_i \in \{0, 1, \dots, L\}\), where \(\delta_i = 0\) indicates censoring and \(\delta_i = l\) indicates the occurrence of event \(l\). We assume that the censoring times \(C_i\) are independent of the event times \((T^*_{i1}, \dots, T^*_{iL})\), i.e., censoring is non-informative.  
Let \(Y_{ijk} = Y_{ik}(s_{ijk})\) denote the value of the \(k\)th longitudinal marker for subject \(i\) at time \(s_{ijk}\). 
As a result, consider 
$\{T_i, \delta_i, Y_{ijk} \mid i = 1, \dots, N; \, j = 1, \dots, n_i; \, k = 1, \dots, K\}$
as a sample from the target population, where \(i\) indexes the subjects in the study, \(j\) indexes the repeated measurements for subject \(i\) with \(n_i\) denoting the total number of longitudinal measurements observed for subject \(i\), and \(k\) indexes the different longitudinal markers.

\subsection{Longitudinal sub-model}
A multivariate linear mixed-effects model is considered for the multiple longitudinal markers as follows:
\begin{eqnarray}\label{m1}
  Y_{ijk} =  \eta_{ik}(s_{ijk}|\bm{\beta}_k,\bm{b}_{ik})+\varepsilon_{ijk}, i=1,~\ldots,N,~j=1,\ldots,n_i,~k=1,\ldots,K,
\end{eqnarray}
where $\varepsilon_{ijk}\overset{\text{iid}}{\sim} N(0, \sigma_k^2)$. We assume that 
\begin{eqnarray}\label{m2}
  \eta_{ik}(s|\bm{\beta}_k,\bm{b}_{ik}) = \bm{X}_{ik}(s)^\top \bm{\beta}_k + \bm{Z}_{ik}(s)^\top \bm{b}_{ik},~k=1,\ldots,K,
\end{eqnarray}
where $\bm{X}_{ik}(s)$ and $\bm{Z}_{ik}(s)$ are the design vectors at time $s$ for the $p^\beta_k$-dimensional fixed effects $\bm{\beta}_k$ and $q_k$-dimensional random effects $\bm{b}_{ik}$, respectively. We assume that $\bm{b}_i=(\bm{b}_{i1}^\top,\ldots,\bm{b}_{iK}^\top)^\top$ takes into account the association within and between the multiple longitudinal markers such that $\bm{b}_i \sim N(\bm{0},\bm{\Sigma})$, and $\bm{\Sigma}$ is a $q \times q$ covariance matrix, where $q = \sum_{k=1}^K q_k$. It is assumed that ${\varepsilon}_{ijk}$ is independent of $\bm{b}_{ik}$, $i=1,\ldots,K$.

\subsection{Competing risks  sub-model}
For each of the $L$ causes, a cause-specific proportional hazard model is postulated. The risk of an event is assumed to be a function of the subject-specific linear predictors $\eta_{ik}(t|\bm{\beta}_k,\bm{b}_{ik})$ and  of time-invariant covariates  as follows \citep{rizopoulos2012joint}:
\begin{eqnarray}\label{h}
\lambda_{il}(t|\bm{\omega}_i,\bm{b}_i)=\lambda_{0l}(t)\exp\{\boldsymbol{\gamma}_l^\top \bm{\omega}_{il}+\sum_{k=1}^{K} \alpha_{lk} \eta_{ik}(t|\bm{\beta}_k,\bm{b}_{ik})\},~l=1,2,\cdots,L,
\end{eqnarray}
where $\lambda_{0l}(t)$ denotes the baseline hazard function for the $l$th cause, $\bm{\omega}_{il}$ is a $p^\gamma_l$ vector of  covariates with corresponding regression coefficients $\boldsymbol{\gamma}_l$, and  $\bm{\alpha}_{l}=({\alpha}_{l1},\ldots,{\alpha}_{lK})^\top$  represents the vector of association parameters between the longitudinal markers and cause $l$. There are different approaches for specifying the baseline hazard function in joint modeling \citep{rizopoulos2012joint}. 
The methodology is not restricted to a specific form of the baseline hazard, and we consider a piecewise constant baseline hazard \citep{ibrahim2001bayesian}. To specify a piecewise constant baseline hazard, we construct a finite partition of the time axis as
$0 = \xi_0 < \xi_1 < \cdots < \xi_\mathcal{K} < \infty,$
such that all observed event or censoring times fall within the last interval, i.e., \(T_i \le \xi_\mathcal{K}\) for all \(i=1,\dots,N\). The intervals are defined as \((\xi_{\kappa-1}, \xi_\kappa]\), \(\kappa=1,\dots,\mathcal{K}\), and the baseline hazard is assumed constant within each interval, \(\lambda_{0l}(t) = \lambda_{\kappa l}\) for \(t \in (\xi_{\kappa-1}, \xi_\kappa]\). In practice, \(\xi_\mathcal{K}\) is chosen slightly larger than the maximum observed time, and the number of intervals \(\mathcal{K}\) is selected to balance model flexibility and parameter stability, often based on quantiles of the observed times or information criteria \citep{he2013estimating}.

\subsection{Parameter estimation of the multi-marker joint model}
We define \(\bm{\psi}_k = (\bm{\beta}_k^\top, \sigma_k^2)^\top\), where \(\bm{\beta}_k\) are the fixed-effect coefficients and \(\sigma_k^2\) the residual variance for the \(k\)th longitudinal marker.  
For the competing risks sub-model, 
$\bm{\Upsilon} = (\boldsymbol{\gamma}_1^\top, \dots, \boldsymbol{\gamma}_L^\top, \bm{\alpha}_1^\top, \dots, \bm{\alpha}_L^\top, \lambda_{11}, \dots, \lambda_{\mathcal{K}L})^\top,$ 
where \(\boldsymbol{\gamma}_l\) are regression coefficients, \(\bm{\alpha}_l\) are association parameters, and \(\lambda_{\kappa l}\) are piecewise constant baseline hazards for cause \(l\).
\\
We define the stacked parameter vector for the longitudinal sub-models as
$\bm{\psi} = (\bm{\psi}_1^\top, \dots, \bm{\psi}_K^\top)^\top,$
and the vector of all observed longitudinal data as
$\bm{y} = (\bm{y}_1^\top, \dots, \bm{y}_N^\top)^\top,$
where \(\bm{y}_i = (\bm{y}_{i1}^\top, \dots, \bm{y}_{iK}^\top)^\top\) is the vector of all repeated measurements for subject \(i\), with \(\bm{y}_{ik} = (y_{i1k}, \dots, y_{in_ik})^\top\) denoting the measurements of marker \(k\).  
\\
Furthermore, \(t_i\) denotes the observed event or censoring time for subject \(i\), i.e., the realization of the random variable
$T_i = \min(T_{i1}^*, \dots, T_{iL}^*, C_i),$
and \(\bm{y}_i\) represents the realization of the corresponding random vector of longitudinal outcomes
$\bm{Y}_i = \{Y_{ijk} : j = 1, \dots, n_i; \, k = 1, \dots, K\}.$
Let \(\bm{t} = (t_1, \dots, t_N)^\top\) and \(\bm{\delta} = (\delta_1, \dots, \delta_N)^\top\). The joint likelihood function for models \eqref{m1} and \eqref{h} is

\begin{align}\label{like}
\mathbb{L}(\bm{\psi},\bm{\Upsilon},\bm{\Sigma} \mid \bm{y}, \bm{t}, \bm{\delta}) 
&= \prod_{i=1}^N \int 
\underbrace{\prod_{k=1}^K p(\bm{y}_{ik} \mid \bm{b}_{ik}, \bm{\psi}_k)}_{\text{longitudinal sub-model}} \\
&\quad \times \underbrace{p(t_i, \delta_i \mid \bm{\omega}_i, \bm{b}_i, \bm{\Upsilon})}_{\text{competing risks sub-model}} 
\underbrace{p(\bm{b}_i \mid \bm{\Sigma})}_{\text{random effects}} d\bm{b}_i \nonumber \\
&= \prod_{i=1}^N \int 
\left( \prod_{k=1}^K \prod_{j=1}^{n_i} \phi(y_{ijk}; \eta_{ik}(s_{ijk} \mid \bm{\beta}_k, \bm{b}_{ik}), \sigma_k^2) \right) \nonumber \\
&\quad \times \left( \prod_{l=1}^L \lambda_{il}(t_i \mid \bm{\omega}_i, \bm{b}_i, \bm{\Upsilon})^{I(\delta_i = l)} 
\exp\left\{ - \sum_{l=1}^L \Lambda_{il}(t_i \mid \bm{\omega}_i, \bm{b}_i, \bm{\Upsilon}) \right\} \right) \nonumber \\
&\quad \times \phi_q(\bm{b}_i; \bm{0}, \bm{\Sigma}) \, d\bm{b}_i, \nonumber
\end{align}
where \(\Lambda_{il}(t \mid \bm{\omega}_i, \bm{b}_i)\) is the cumulative hazard function of Equation \eqref{h}, and \(\phi(x; \mu, \sigma^2)\) and \(\phi_q(\bm{x}; \bm{\mu}, \bm{\Sigma})\) denote the densities of the univariate and \(q\)-variate normal distributions, respectively.  
Also, \(I(\delta_i = l)\) denotes the indicator function, defined as
\[
I(\delta_i = l) =
\begin{cases}
1, & \text{if } \delta_i = l,\\
0, & \text{otherwise.}
\end{cases}
\]
It ensures that the hazard of cause \(l\) only contributes to the likelihood if subject \(i\) experiences event \(l\); otherwise, it contributes 1.
\\
In the frequentist paradigm, parameter estimation can be computed by maximizing \eqref{like}, which involves the use of numerical integration \citep{chi2006joint,song2002estimator}. This integration will be more challenging when the number of outcomes increases. 
In the Bayesian paradigm \citep{xu2001evaluation,rizopoulos2011bayesian,baghfalaki2014joint,long2018joint}, parameter estimation is based on the joint posterior distribution
\begin{align}\label{post}
\pi(\boldsymbol{\psi}, \boldsymbol{\Upsilon}, \boldsymbol{b}, \boldsymbol{\Sigma} \mid \boldsymbol{y}, \boldsymbol{t}, \boldsymbol{\delta}) 
&\propto \prod_{i=1}^N 
\Bigg[ \prod_{k=1}^K \prod_{j=1}^{n_i} p\left(y_{ijk} \mid \boldsymbol{b}_{ik}, \boldsymbol{\psi}_k\right) \, p(\boldsymbol{\psi}_k) \Bigg] \\
&\quad \times \prod_{l=1}^L \lambda_{il}(t_i \mid \bm{\omega}_i, \bm{b}_i, \bm{\Upsilon})^{I(\delta_i=l)} 
\exp\Big\{- \sum_{l=1}^L \Lambda_{il}(t_i \mid \bm{\omega}_i, \bm{b}_i, \bm{\Upsilon}) \Big\} \nonumber \\
&\quad \times \phi_q(\boldsymbol{b}_i ; \mathbf{0}, \boldsymbol{\Sigma}) \, p(\boldsymbol{\Upsilon}) \, p(\boldsymbol{\Sigma}), \nonumber
\end{align}
where \(\boldsymbol{b} = (\boldsymbol{b}_1^\top, \dots, \boldsymbol{b}_N^\top)^\top\) is the stacked vector of random effects for all subjects, with \(\boldsymbol{b}_i = (\boldsymbol{b}_{i1}^\top, \dots, \boldsymbol{b}_{iK}^\top)^\top\) representing the random effects for the \(K\) longitudinal markers of subject \(i\). 
The prior distributions are chosen to be weakly informative to allow the data to dominate the posterior while ensuring computational stability: 
 Regression coefficients \(\boldsymbol{\beta}_k\) and association parameters \(\boldsymbol{\alpha}_l\): zero-mean normal distributions with large variance; 
   Residual variances \(\sigma_k^2\): inverse gamma distributions with small shape and scale parameters; 
 Random effects covariance \(\boldsymbol{\Sigma}\): inverse Wishart distribution with degrees of freedom equal to the dimension of \(\boldsymbol{b}_i\) and identity scale matrix; 
 Baseline hazard parameters \(\lambda_{\kappa l}\): gamma distributions with mean 1 and large variance.
 \\
These choices provide sufficient flexibility while maintaining numerical stability, avoiding the potential convergence and identifiability issues associated with fully non-informative priors \citep{baghfalaki2014joint,baghfalaki2021approximate}.\\
In the next section, we introduce our proposed  two-stage approach  for variable selection of longitudinal markers and covariates in a Bayesian paradigm.

\subsection{Two-stage joint modeling of multiple longitudinal measurements and competing risks}
The proposed method adopts a two-stage approach for variable selection in joint modeling, targeting the selection of markers (or time-varying covariates) and time-invariant covariates. This approach is specifically designed for handling multiple longitudinal markers and time-to-event outcomes, with the primary goal of avoiding bias caused by informative dropouts.
This relies on the idea of TSJM approach of \cite{Baghfalakitsjm} but considering competing risks and replacing the estimation of a Cox model by partial likelihood in the second stage by a fully Bayesian approach allowing variable selection.\\
The first stage involves estimating $K$ one-marker joint models for the competing risks and each marker. In the second stage, we estimate cause-specific proportional hazard models including the time-invariant explanatory variables and the predicted current values of the linear predictor of each longitudinal markers obtained from stage 1. Therefore, the approach consists of two stages, outlined as follows:
\begin{description}
\item[Stage 1] For each $k$, $k=1,\ldots,K$, estimate the joint model for the longitudinal marker $k$ and competing risks. The one-marker JMs are estimated using the joint posterior distribution given by Equation \eqref{post} for $K=1$. 
    Then, predict the random effects $\hat{\bm{b}}_{ik}$ and the linear predictor $\eta_{ik}(t|\hat{\bm{\beta}}_k,\hat{\bm{b}}_{ik})$, for marker $k$ using the empirical mean of the posterior distribution $(\bm{b}_{ik}|\bm{y}_{ik},t_i,\delta_i,\hat{\bm{\theta}}_k)$, where $\hat{\bm{\theta}}_k$ represents the estimated parameters for the one-marker joint model for the $k$th longitudinal marker. 
This first stage allow the estimation of $\eta_{ik}(t|\hat{\bm{\beta}}_k,\hat{\bm{b}}_{ik})$ without the bias due to informative dropout but neglecting the correlation between the markers. 
  \item[Stage 2] In the second stage, the values of $\eta_{ik}(t|\hat{\bm{\beta}}_k,\hat{\bm{b}}_{ik})$ are included as time-varying covariates in a proportional hazard model to estimate parameters $\boldsymbol{\gamma}$ and $\bm{\alpha}_k$, as well as the baseline hazard from \eqref{h}.
  The cumulative hazard function of \eqref{h} can be approximated using Gaussian quadrature \citep{bogaert2014iteration,tanton2005encyclopedia}.  In the Bayesian paradigm, the following joint posterior distribution is considered:
\begin{eqnarray}\label{hazz2}
\pi(\bm{\lambda}, \boldsymbol{\alpha} \mid \boldsymbol{t}, \boldsymbol{\delta}, \hat{\boldsymbol{b}})  & \propto & \prod_{i=1}^N \prod_{l=1}^L \bigr[\lambda_{il}(t_i|\bm{\omega}_i,\hat{\bm{b}}_i,\bm{\Upsilon}_l) \bigr]^{I(\delta_i=l)}\\\nonumber &\times& \exp\bigr(-\sum_{l=1}^L  \Lambda_{il}(t_i|\bm{\omega}_i,\hat{\bm{b}}_i,\bm{\Upsilon}_l)  \bigr)
 \times p(\boldsymbol{\Upsilon}),
\end{eqnarray}
 where  $\boldsymbol{\Upsilon}_l = (\boldsymbol{\gamma}_l^\top, \boldsymbol{\alpha}_l^\top, \boldsymbol{\lambda}_l^\top)^\top$,
with 
$\boldsymbol{\lambda}_l = (\lambda_{l1}, \ldots, \lambda_{l\mathcal{K}})^\top$
denoting the vector of piecewise-constant baseline hazards for cause \(l\).  
We also define 
$\boldsymbol{\alpha} = (\boldsymbol{\alpha}_1^\top, \ldots, \boldsymbol{\alpha}_L^\top)^\top$   and  
$\boldsymbol{\lambda} = (\lambda_{11}, \ldots, \lambda_{\mathcal{K}L})^\top.$  
Furthermore, \(\boldsymbol{\Upsilon}\) is as defined in Equation~\eqref{like}, and the prior distributions for all unknown parameters are the same as those specified for Equation~\eqref{post}.
 \end{description}

\subsection{Bayesian variable selection}
As above, we consider the prior distributions
$\boldsymbol{\beta}_k \sim N_{p_k}(\boldsymbol{0}, c \, \mathbf{I}_{p_k})$,
where \(\boldsymbol{0}\) is a \(p_k\)-dimensional vector of zeros, \(\mathbf{I}_{p_k}\) is the \(p_k \times p_k\) identity matrix, and \(c\) is a large positive constant,
$\sigma_k^2\sim {\rm{I\Gamma}}(a_\sigma,b_\sigma)$, $\bm{\Sigma}\sim \rm{IWishart}(\bm{\mathcal{V}},\nu)$ and $\lambda_{l\kappa}\sim {\rm{\Gamma}}(a_{l\kappa},b_{l\kappa}),~{\kappa}=1,\ldots,\mathcal{K},~l=1,\ldots,L$. Our purpose is to implement variable selection for the regression coefficients $\boldsymbol{\gamma}_l,~l=1,\ldots,L$ of the cause-specific hazard model \eqref{h} and the association parameters  $\bm{\alpha}_{l},~l=1,\ldots,L$ of \eqref{h}. Therefore, we consider two different prior distributions, including continuous spike and Dirac spike, for them. \\
A key advantage of the Bayesian variable selection approach is that the posterior distributions of the coefficients $\boldsymbol{\gamma}_l$ and $\boldsymbol{\alpha}_l$ naturally account for both parameter uncertainty and the uncertainty arising from the variable selection process itself \cite{George1993,O'Hara2009,Malsiner-Walli2018}. Consequently, the posterior means and standard deviations provide valid measures of effect size and uncertainty without the need to re-estimate the model using only the selected variables, as is commonly done in frequentist post-selection procedures. This property ensures that credible intervals and inference fully reflect the selection process.

\subsubsection{Continuous spike}
\begin{figure}[ht]
\centering
\includegraphics[width=9cm]{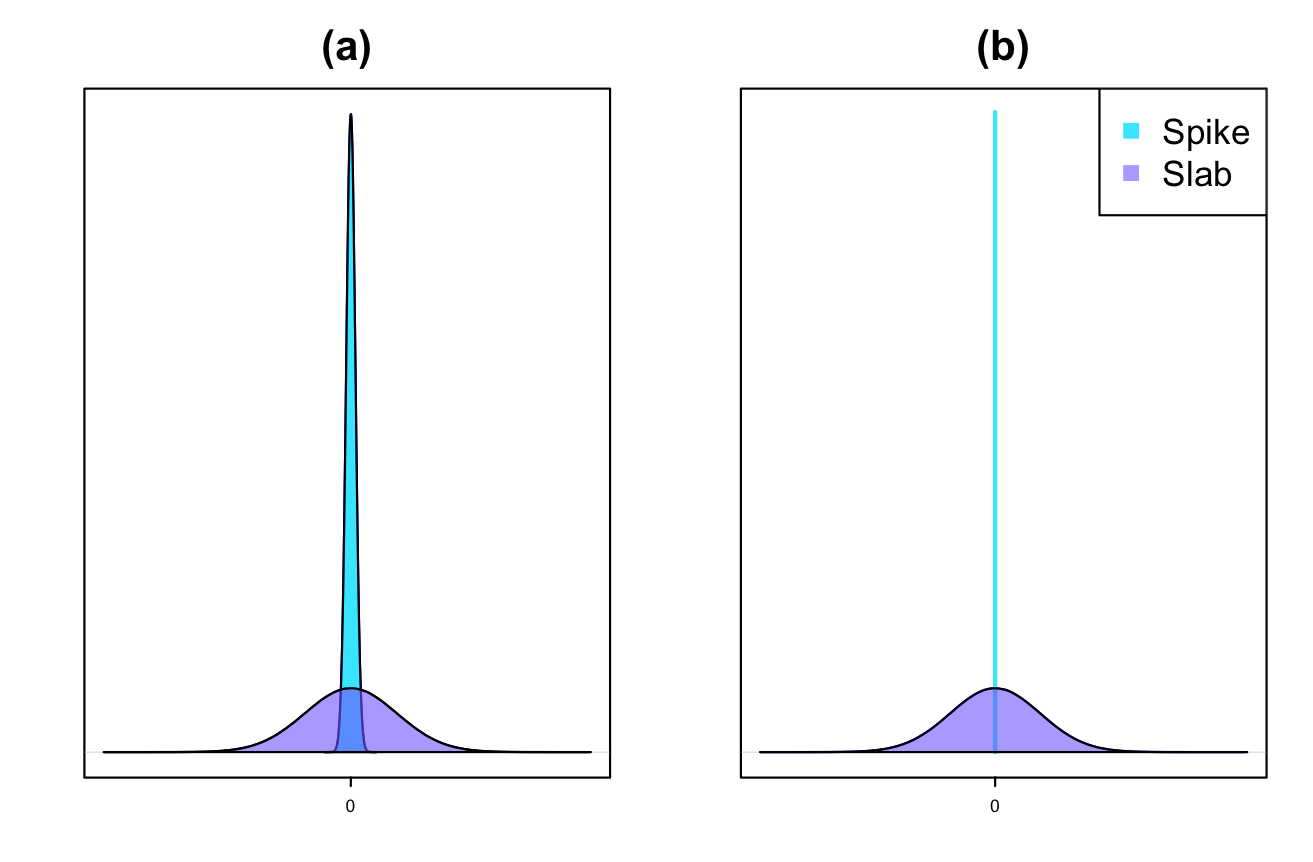}
\vspace*{-.1cm} \caption{\label{csds}
An example of continuous spike prior (Panel a) and Dirac spike prior (Panel b).
 }
\end{figure}

The continuous spike \citep[CS,][]{malsiner2018comparing,george1993variable}  prior for $\boldsymbol{\gamma}_l=(\gamma_{l1},\ldots,\alpha_{p_l^{\gamma_{l}}})^\top$ is given by
\begin{eqnarray}\label{csp2}
\begin{array}{l}
{\gamma_{lj}}|{\zeta_{lj}},\sigma_{\gamma_{lj}}^2 ,\tau_{\gamma_{lj}}^2 \mathop{\sim}\limits^{ind} (1 - {\zeta_{lj}}){N}({0},\sigma_{\gamma_{lj}}^2) + {\zeta_{lj}}{N}({0},\tau_{\gamma_{lj}}^2 ), ~~~~\sigma_{\gamma_{lj}}^2 > \tau_{\gamma_{lj}}^2 > 0, ~l=1,\ldots,L,~j=1,\ldots,p^\gamma_l,\\
{\zeta_{lj}}|\pi_{\gamma_{lj}} \mathop{\sim}\limits^{ind}  Ber(\pi_{\gamma_{lj}}),\\
\pi_{\gamma_{lj}} |{a_{\gamma_{lj}}},{b_{\gamma_{lj}}}\mathop{\sim}\limits^{ind} Beta(a_{\gamma_{lj}},b_{\gamma_{lj}}),\\
\sigma_{\gamma_{jk}}^2 | a_{\sigma_{\gamma_{jk}}^2},b_{\sigma_{\gamma_{jk}}^2}\mathop{\sim}\limits^{ind} {\rm{I\Gamma}}(a_{\sigma_{\gamma_{jk}}^2},b_{\sigma_{\gamma_{jk}}^2}).
\end{array}
\end{eqnarray}
      Also, the CS prior for the association parameter  $\bm{\alpha}_l=(\alpha_{l1},\ldots,\alpha_{lK})^\top$ is given by
\begin{eqnarray}\label{csp3}
\begin{array}{l}
{\alpha_{lk}}|{\zeta_{lk}},\sigma_{\alpha_{lk}}^2 ,\tau_{\alpha_{lk}}^2 \mathop{\sim}\limits^{ind} (1 - {\zeta_{lk}}) N(0,\sigma_{\alpha_{lk}}^2) + {\zeta_{lk}} N(0,\tau_{\alpha_{lk}}^2), ~~~~ \sigma_{\alpha_{lk}}^2 > \tau_{\alpha_{lk}}^2 > 0, ~k=1,\ldots,K,\\
{\zeta_{lk}}|\pi_{\alpha_{lk}} \mathop{\sim}\limits^{ind} Ber(\pi_{\alpha_{lk}}),\\
\pi_{\alpha_{lk}} | a_{\alpha_{lk}}, b_{\alpha_{lk}}\mathop{\sim}\limits^{ind} Beta(a_{\alpha_{lk}}, b_{\alpha_{lk}}),\\
\sigma_{\alpha_{lk}}^2 | a_{\sigma_{\alpha_{lk}}^2}, b_{\sigma_{\alpha_{lk}}^2}\mathop{\sim}\limits^{ind} {\rm{I\Gamma}}(a_{\sigma_{\alpha_{lk}}^2}, b_{\sigma_{\alpha_{lk}}^2}).
\end{array}
\end{eqnarray}
where, $\tau_{\gamma_{lj}}^2$ in \eqref{csp2} and $\tau_{\alpha_{lk}}^2$ in \eqref{csp3} are a small fixed value (e.g., $10^{-3}$ or $10^{-4}$) and represents a spike prior. The latent variable ${\zeta_{lj}}$ in \eqref{csp2} (or $\zeta_{lk}$ in \eqref{csp3}) is considered to account for the significance of the parameter. For example, if ${\zeta_{jk}}=0$, then the distribution of ${\zeta_{lk}},\sigma_{\alpha_{lk}}^2 ,\tau_{\alpha_{lk}}^2$ will be ${N}({0},\sigma_{\alpha_{lk}}^2)$, and  can be considered as significant. This leads us to have a slab prior. A graphical example of CS prior can be found in panel (a) of Figure \ref{csds}.

\FloatBarrier
\subsubsection{Dirac spike}
We exploit a Dirac spike-and-slab prior \citep[DS,][]{george1993variable,mitchell1988bayesian} similar to that proposed by \cite{xu2015bayesian}.
For this purpose, the DS prior for $\boldsymbol{\gamma}_l$ is given by
\begin{eqnarray}\label{dsp2}
\begin{array}{l}
{\gamma_{lj}}|\pi_{\gamma_{lj}},\sigma_{\gamma_{lj}}^2 \mathop{\sim}\limits^{ind} (1 - \pi_{\gamma_{lj}}) N(0,\sigma_{\gamma_{lj}}^2) + \pi_{\gamma_{lj}} \delta_0(\gamma_{lj}), ~\sigma_{\gamma_{lj}}^2 > 0, ~l=1,\ldots,L,~j=1,\ldots,p^\gamma_l,\\
\pi_{\gamma_{lj}} | a_{\gamma_{lj}}, b_{\gamma_{lj}}\mathop{\sim}\limits^{ind} Beta(a_{\gamma_{lj}}, b_{\gamma_{lj}}),\\
\sigma_{\gamma_{lj}}^2 | a_{\sigma_{\gamma_{lj}}^2}, b_{\sigma_{\gamma_{lj}}^2}\mathop{\sim}\limits^{ind} {\rm{I\Gamma}}(a_{\sigma_{\gamma_{lj}}^2}, b_{\sigma_{\gamma_{lj}}^2}).
\end{array}
\end{eqnarray}
where $\delta_0(\gamma_{lj})$ represents a point mass at $0 \in \mathbb{R}$. It is defined as $\delta_0(\gamma _{lj})=1$ when $\gamma _{lj}={0}$ and $\delta_0(\gamma _{lj})=0$ when $\gamma _{lj}$ is non-zero, i.e., $\gamma _{lj}\neq {0}$.
      Also, the DS prior for the association parameter  $\bm{\alpha}$ is given by
\begin{eqnarray}\label{dsp3}
\begin{array}{l}
{\alpha_{lk}}|\pi_{\alpha_{lk}}, \sigma_{\alpha_{lk}}^2 \mathop{\sim}\limits^{ind} (1 - \pi_{\alpha_{lk}}) N(0,\sigma_{\alpha_{lk}}^2) + \pi_{\alpha_{lk}} \delta_0(\alpha_{lk}), ~ \sigma_{\alpha_{lk}}^2 > 0, ~k=1,\ldots,K,\\
\pi_{\alpha_{lk}} | a_{\alpha_{lk}}, b_{\alpha_{lk}}\mathop{\sim}\limits^{ind} Beta(a_{\alpha_{lk}}, b_{\alpha_{lk}}),\\
\sigma_{\alpha_{lk}}^2 | a_{\sigma_{\alpha_{lk}}^2}, b_{\sigma_{\alpha_{lk}}^2}\mathop{\sim}\limits^{ind} {\rm{I\Gamma}}(a_{\sigma_{\alpha_{lk}}^2}, b_{\sigma_{\alpha_{lk}}^2}).
\end{array}
\end{eqnarray}
A graphical example of the DS prior can be found in panel (b) of Figure \ref{csds}. For implementing Bayesian inference, a block Gibbs sampler is used in conjunction with the Metropolis-Hastings algorithm. For this purpose, we require all of the full conditional distributions. The Markov Chain Monte Carlo (MCMC) sampling is performed using ``JAGS" \citep{plummer2003jags,plummer2012jags} and the \texttt{R2jags} \citep{su2015r2jags} package is utilized as an interface between R and \texttt{JAGS}. The package automatically computes the full conditional posterior distributions for all parameters and models. 
We refer to this approach as Variable Selection for Joint Modeling (VSJM).
The methodology has been implemented in the R package \texttt{VSJM}.
\subsection{Statistical inference }
To select significant variables in the covariates of the cause-specific hazard model and by considering Equation \eqref{csp2} or \eqref{dsp2}, the following hypothesis tests are conducted:
\begin{eqnarray}\label{h1}
\left\{ {\begin{array}{*{20}{c}}
   {{{\rm{H}}_{0lj}^\gamma}:{\gamma_{lj}} = 0}  \\
   {{{\rm{H}}_{1lj}^\gamma}:{\gamma_{lj}} \ne 0}  \\
\end{array}} \right.,~~j=1,\ldots,p_l^\gamma,l=1,\ldots,L,
 \end{eqnarray}
Also, to select significant longitudinal markers based on the association parameters and by considering Equation \eqref{csp3} or \eqref{dsp3}, the following hypothesis tests are considered:
 \begin{eqnarray}\label{h2}
\left\{ {\begin{array}{*{20}{c}}
   {{{\rm{H}}_{0lk}^\alpha}:{\alpha _{lk}} = 0}  \\
   {{{\rm{H}}_{1lk}^\alpha}:{\alpha _{lk}} \ne 0}  \\
\end{array}} \right.,~~k=1,\ldots,K,l=1,\ldots,L,.
 \end{eqnarray}
In practice, when implementing the hypotheses \eqref{h1} and \eqref{h2} for CS, it is possible to consider the following equivalent hypotheses, respectively:
\begin{eqnarray}\label{ht2}
\left\{ {\begin{array}{*{20}{c}}
   {{{\rm{H}}_{0lj}^\gamma}:{\zeta_{lj}} = 0}  \\
   {{{\rm{H}}_{1lj}^\gamma}:{\zeta_{lj}} \ne 0}  \\
\end{array}} \right.,~~j=1,\ldots,p_l^\gamma,l=1,\ldots,L,
 \end{eqnarray}
 \begin{eqnarray}\label{ht3}
\left\{ {\begin{array}{*{20}{c}}
   {{{\rm{H}}_{0lk}^\alpha}:{\zeta _{lk}} = 0}  \\
   {{{\rm{H}}_{1lk}^\alpha}:{\zeta _{lk}} \ne 0}  \\
\end{array}} \right.,~~k=1,\ldots,K,l=1,\ldots,L,.
 \end{eqnarray}
We consider two different Bayesian testing approaches for evaluating these hypotheses: a local Bayesian false discovery rate \citep{efron2012large} and a Bayes factor \citep{kass1995bayes}.
\subsubsection{Continuous spike}
{\bf{Local Bayesian false discovery rate}}\\
Let $\bm{\cal{D}}$ denote the data in the study. The set of unknown parameters of the CS model is denoted as $\bm{\theta}$. Let $P({\rm{{\rm{H}}}}_{0k}|\bm{\cal{D}})$ represent the posterior probability of the null hypothesis. This is the probability of making a false discovery for a significant covariate, which is called the local Bayesian false discovery rate for $\bm{\cal{D}}$
(denoted by $lBFDR$) by  Efron \citep{efron2012large}. The small values of $lBFDR$ indicate strong evidence for the presence of a significant covariate. For computing $lBFDR$ of Equation \eqref{ht2}, we obtain
\begin{eqnarray*}
lBFDR_{lj}^\gamma &=& P({{\rm{{\rm{H}}}}^\gamma_{0lj}}|\bm{\cal{D}}) = P({\zeta_{lj}} = 0|\bm{\cal{D}})\\
 &\approx& \frac{1}{M}\sum\limits_{m = 1}^M {I({\zeta_{lj}} = 0|\bm{\cal{D}},\bm{\theta}^{(m)} )},
  \end{eqnarray*}
   where $I({\zeta_{lj}} = 0|\bm{\cal{D}},\bm{\theta}^{(m)} )$ is an indicator function and 
   $\bm{\theta}^{(m)}$ denotes  the  $m^{th}, m=1,\ldots,M$, generated sample of
$\bm{\theta}$ using the Gibbs sampler.\\
The $lBFDR$ is a procedure for investigating the null hypothesis. It is an empirical Bayes version of the methodology proposed by \cite{benjamini1995controlling}. This approach focuses on densities instead of tail areas \citep{efron2001scales,efron2002empirical,efron2004large}. Efron \citep{efron2004large} demonstrated that the false discovery rate control rule proposed by \cite{benjamini1995controlling} can be expressed in terms of empirical Bayes. They also showed that the desired control level (referred to as $\alpha[=0.05]$) for Benjamini-Hochberg testing of single tests can be regarded as the desired control level for local Bayes false discovery rate ($lBFDR$) as well. Based on Remark B of \cite{efron2004large}, if the value of $lBFDR$ is not greater than 0.05 (the desired control level), the null hypothesis will be rejected. In fact, the interpretation of the $lBFDR$ value is analogous to the frequentist's p-value. $lBFDR$ values less than a specified level (e.g., 0.05) of significance provide stronger evidence against the null hypothesis \citep{sampson2013controlling,yap2023bayesian}. The value of 0.05 in the context of variable selection is also considered a threshold for $lBFDR$ by \cite{majumdar2018efficient} and \cite{baghfalaki2021bayesian}.
Thus, a local Bayesian false discovery rate of 0.05 is considered in this paper.\\
{\bf{Bayes factor}}\\
The Bayes factor for testing \eqref{ht2} is defined as
\begin{eqnarray*}
BF^\gamma_{lj} &=& \frac{{P({{\rm{{\rm{H}}}}_{1lj}^\gamma}|\bm{\cal{D}})/P({{\rm{{\rm{H}}}}_{1lj}^\gamma})}}{{P({{\rm{{\rm{H}}}}_{0lj}^\gamma}|\bm{\cal{D}})/P({{\rm{{\rm{H}}}}_{0lj}^\gamma})}}\\
 &=& \frac{{P({\zeta_{lj}} \ne 0|\bm{\cal{D}})/P({\zeta_{lj}} \ne 0)}}{{P({\zeta_{lj}} = 0|\bm{\cal{D}})/P({\zeta_{lj}} =  0)}},
  \end{eqnarray*}
which describes the evidence of  ${\rm{{\rm{H}}}}_{1k}$ against ${\rm{{\rm{H}}}}_{0k}$. Note that
$P({\zeta_{lj}} = 1) = E\left[ \pi_{\gamma_{lj}}  \right] = 
\frac{{a_{\gamma_{lj}}}}{{a_{\gamma_{lj}}}+{b_{\gamma_{lj}}}}$ as
$\zeta_{11},\ldots,\zeta_{L p^\gamma_L}$ are assumed independent. Hence,
\begin{eqnarray*}
P(\zeta_{lj} = 0)   = \frac{b_{\gamma _{lj}}}{a_{\gamma _{lj}}+{b_{\gamma _{lj}}}}.
   \end{eqnarray*}
Also, since $lBFDR_{lj}^\gamma = P({{\rm{{\rm{H}}}}_{0lj}^\gamma}|\bm{\cal{D}})$,
\begin{eqnarray}\label{BF}
BF_{lj}^\gamma = \frac{{1 - lBFDR_{lj}^\gamma}}{{lBFDR_{lj}^\gamma}} \times \frac{b_{\gamma _{lj}}}{a_{\gamma _{lj}}}.
   \end{eqnarray}
Unlike $lBFDR$, a large value of $BF$ indicates strong evidence in favor of ${\rm{{\rm{H}}}}_{1}$. Also, a widely cited table is provided by \cite{kass1995bayes}. This table is based on $log_{10}(BF_k)$ and can be used to quantify the support that one model has over another.\\
All comments mentioned above can also be applied to the hypothesis testing of Equations \eqref{ht3}.
 
\subsubsection{Dirac spike}
The strategy for performing $lBFDR$ and $BF$ for DS is to define indicator variables as $\Psi^\gamma_{lj}$, $j=1,\ldots,p_l^\gamma,l=1,\ldots,L,$  and $\Psi^\alpha_{lk}$, $k=1,\ldots,K$ such that
 $${\Psi^\gamma_{lj}} = \left\{ {\begin{array}{*{20}{c}}
1 & {\rm{if} ~~{{\gamma_{lj}} \ne {0}}}\\
0 & {\rm{if} ~~{{\gamma_{lj}} = {0}}}
\end{array}}\right.,$$
 $${\Psi^\alpha _{lk}} = \left\{ {\begin{array}{*{20}{c}}
1&{\rm{if} ~~ {\alpha _{lk}} \ne {0}}\\
0&{\rm{if} ~~{\alpha _{lk}} = {0}}
\end{array}}\right..$$
By considering these indicator variables, all of those described for the CS can also be used for the DS.

\section{Dynamic predictions}
\subsection{Dynamic predictions from VSJM}
The joint modeling of longitudinal measurements and time-to-event outcomes is a popular methodology for calculating dynamic prediction (DP). The DP represents the risk predictions for subject $i$ who has survived up to time point $s$, taking into account longitudinal measurements and other information available until time $s$. Consider the notation described in Section 2.
Let the proposed joint models be fitted to a random sample, which is referred to as training data. For validation purposes, we are interested in predicting the risk of subjects in a new set of data called the validation set.
The longitudinal measurements until time $s$ are denoted by $\bm{\mathcal{Y}}_{i}(s)$ and given by $\{Y_{ik}(s_{ijk}); 0\le s_{ijk}\le s, j=1,\ldots,n_i, k=1,\ldots,K \}$. Additionally, the vector of available explanatory variables until time $s$ is $\bm{\mathcal{X}}_{i}(s)$ and given by $\{\bm{w}_i, \bm{X}_{ik}(s_{ijk}); 0\le s_{ijk}\le s, j=1,\ldots,n_i, k=1,\ldots,K \}$.
Let $\delta_i=1$ denote the event of interest, then, for the landmark time $s$ and the prediction window $t$, the risk probability is  as follows:
\begin{eqnarray}\label{dp}
\pi_i(s+t \mid s) & = &\mathrm{P}\left(s \leq T_i^*<s+t,\delta_i=1 \mid T_i^*>s, \mathcal{Y}_i(s), \boldsymbol{\mathcal { X }}_i(s) ; \boldsymbol{\theta}\right),
\end{eqnarray}
where ${\bm{\theta}}$ is a vector of all unknown parameters of the model, including ${\bm{\theta}}_k$ which is estimated at stage 1, and ${\boldsymbol{\Upsilon}}$ which is estimated at stage 2, and $T_i^*$ represents the true event time for the $i$th subject.
The predicted probability \eqref{dp} can be computed as:
\begin{eqnarray}\label{dppp}
\begin{aligned}
&\mathrm{P}\Big(s \leq T_i^*<s+t, \delta_i=1 \mid T_i^*>s, \mathcal{Y}_i(s), \boldsymbol{\mathcal { X }}_i(s) ; \boldsymbol{\theta}\Big) \\
&= \int 
\frac{\mathrm{P}\Big(s \leq T_i^*<s+t, \delta_i=1 \mid \boldsymbol{b}_i, \boldsymbol{\mathcal { X }}_i(s) ; \boldsymbol{\theta}\Big)}
{S\Big(s \mid \boldsymbol{b}_i, \boldsymbol{\mathcal { X }}_i(s) ; \boldsymbol{\theta}\Big)} \\
& \quad \times f\Big(\boldsymbol{b}_i \mid T_i^*>s, \mathcal{Y}_i(s), \boldsymbol{\mathcal { X }}_i(s) ; \boldsymbol{\theta}\Big) \, d \boldsymbol{b}_i
\end{aligned}
\end{eqnarray}
where $S(\cdot)$ represents the
survival probability to survive to both events given random effects.
 The first order approximation for estimating this quantity \citep{rizopoulos2011dynamic,Baghfalakitsjm} is given as follows:
\begin{eqnarray*}\label{dpcr2}
\hat\pi_i(s+t \mid s)  =\frac{\int_{s}^{s+t} 
 \exp\{-\sum_{l=1}^{L} \Lambda_l(u\mid  \hat{\bm{b}}_{i},\bm{\mathcal{X}}_{i}(s); \hat{\bm{\theta}})\}
 \lambda_1(u\mid  \hat{\bm{b}}_{i}, \bm{\mathcal{X}}_{i}(s); \hat{\bm{\theta}})\}du}{\exp\{-\sum_{l=1}^{L} \Lambda_l(s\mid  \hat{\bm{b}}_{i},\bm{\mathcal{X}}_{i}(s); \hat{\bm{\theta}})\}},
\end{eqnarray*}
where $\hat{\bm{\theta}}=\{\hat{\bm{\theta}}_1,\ldots,\hat{\bm{\theta}}_K, \hat{\boldsymbol{\Upsilon}}\}$ are estimated based on the training data.
To predict $\hat{\bm{b}}_{i}$,   we consider the one-marker joint models for each marker and predict the $p_k$-dimensional random effect $\bm{b}_{ik}$ given $(T^*_i > s,\bm{\mathcal{Y}}_{i}(s),\bm{\mathcal{X}}_{i}(s), \hat{\bm{\theta}}_k)$. This is denoted by $\hat{\bm{b}}_{ik}$, where $\hat{\bm{\theta}}_k$ represents the estimated parameters for the one-marker joint model in the first stage of VSJM. Merge the predicted random effects to obtain the $p=\sum_{k=1}^Kp_k$-dimensional predicted random effect $\hat{\bm{b}}_{i}=(\hat{\bm{b}}_{i1},\ldots,\hat{\bm{b}}_{iK})$.\\ 
For Monte-Carlo approximation of \eqref{dppp}, we follow the same strategy as those discussed in \cite{zhou2010note}. For this aim, 
consider a realization of the generated MCMC sample for estimating the unknown parameters from the one-marker JM for marker $k$ in the training data, denoted $\bm{\theta}_k^{(\ell)},~\ell=1,\ldots,\mathcal{L}$.
Draw $M$ values $\bm{b}_{ik}^{(\ell,m)}$ from the posterior distribution $(\bm{b}_{ik} | T^*_i > s,\bm{\mathcal{Y}}_{i}(s),\bm{\mathcal{X}}_{i}(s), \bm{\theta}_k^{(\ell)})$. For $~\ell=1,\ldots,\mathcal{L},~m=1,\ldots,M,$
    \begin{eqnarray*}
\hat\pi_{i}(s+t|s)^{(\ell,m)} =
\frac{\int_{s}^{s+t} 
 \exp\{-\sum_{l=1}^{L} \Lambda_l(u\mid  {\bm{b}}_i^{(\ell,m)},\bm{\mathcal{X}}_{i}(s); \bm{\theta}_k^{(\ell)})\}
 \lambda_1(u\mid  {\bm{b}}_i^{(\ell,m)}, \bm{\mathcal{X}}_{i}(s); \bm{\theta}_k^{(\ell)})\}du}{\exp\{-\sum_{l=1}^{L} \Lambda_l(s\mid  {\bm{b}}_i^{(\ell,m)},\bm{\mathcal{X}}_{i}(s); \bm{\theta}_k^{(\ell)})\}}.
\end{eqnarray*}
Calculate any desired quantity from the generated chains, such as percentiles, to determine the credible interval.

\subsection{Reducing bias in dynamic predictions from VSJM}
To reduce estimation biases resulting from variable selection, we propose incorporating an additional stage to calculate dynamic predictions. After variable selection using CS or DS prior, we recommend re-estimating the proportional hazard model by substituting CS or DS with non-informative normal priors for the association parameters of the selected markers and the regression coefficients of the selected covariates. Then calculate the DP as described in Section 3.1.

\subsection{Measures of prediction accuracy}
We use the Area Under the Receiver Operating Characteristics Curve (AUC) and the Brier Score (BS) to assess the predictive performance of prediction models \citep{steyerberg2009applications}.
AUC represents the probability that the model correctly ranks a randomly chosen individual with the disease higher than a randomly chosen individual without the disease based on their predicted risks
Additionally, BS  measures the preduction accuracy by quantifying the mean squared error of prediction.
For censored and competing time-to-event and time-dependent markers, a non-parametric inverse probability of censoring weighted estimator of AUC and BS was proposed by \cite{blanche2015quantifying} and utilized in this paper.

\section{Simulation study}
\subsection{Design of simulation study}
A simulation study was conducted to assess the effectiveness of the proposed VSJM approach for variable selection and risk prediction. To evaluate the effectiveness of the proposed approaches for variable selection, we examine three distinct criteria: the true positive rate (TPR), which measures the proportion of correctly selected variables among all significant  variables; the false positive rate (FPR), which measures the proportion of incorrectly selected variables among all irrelevant  variables; and the Matthews correlation coefficient 
 \citep{MATTHEWS1975442} (MCC).
The latter is defined as follows:
 \begin{eqnarray}
 \rm{MCC} &=&\frac{\rm{TP}\times \rm{TN} -\rm{FP}\times \rm{FN}}{\sqrt{(\rm{TP+FP)(TP+FN)( TN+FP)(TN+FN)}}},
 \end{eqnarray}
where TP, TN, FP, and FN represent the number of true positives, true negatives, false positives, and false negatives, respectively. The MCC and TPR are expected to reach 1, while the FPR is expected to be close to zero for optimal performance. $BF$ and $lBFDR$ are used for variable selection, with the results based on a threshold of 1 for $BF$ and a threshold of 0.05 for $lBFDR$.\\
The simulation scenarios varied based on the strength of the association between the markers and the events, the residual variances for the markers that quantify measurement error, and the correlation between the markers that is neglected in the first stage of the approach. The performance of the approaches is also evaluated based on their predictive abilities using AUC and BS. The predictive  performance of the approach is compared with that of the true model using the true values of the parameters and random effects (referred to as ``Real") as well as the oracle model (true model, where parameters are estimated by accounting for the multi-marker joint modeling).\\
Data were generated using multi-marker joint models with $K=10$ longitudinal markers and a competing risks model with two causes. The longitudinal outcomes were simulated from the following model:
\begin{align}\label{gau}
Y_{ijk} &= \eta_{ik}(s_{ijk}\mid \bm{\beta}_k, \bm{b}_{ik}) + \varepsilon_{ijk} \\
        &= \beta_{0k} + \beta_{1k} s_{ijk} + b_{0ik} + b_{1ik} s_{ijk} + \varepsilon_{ijk}, \nonumber
\end{align}
where $\varepsilon_{ijk} \sim N(0, \sigma_k^2)$, and 
\[
\bm{b}_i = (b_{0i1}, b_{1i1}, b_{0i2}, b_{1i2}, \ldots, b_{0i,10}, b_{1i,10})^{\top} \sim N(\bm{0}, \bm{\Sigma}), \qquad i=1,\ldots,N.
\]
For each subject $i$, marker $k = 1,\ldots,10$, and measurement index $j$, the observation times were defined as
$s_{ijk} \in \{0,\, 0.1,\, 0.2,\, \ldots,\, 2\}$,
corresponding to equally spaced measurements over the interval $[0, 2]$. The fixed-effect coefficients were set to
$\bm{\beta}_k = (-0.5,\, 0.5)^{\top},  ~ k = 1,\ldots,10$.\\
 The time-to-event outcome was simulated using a cause-specific hazard model that depends on the current value of the $K$ markers and on 24 time-invariant covariates as follows:
\begin{eqnarray}\label{hazsim}
\lambda_l(t)=\lambda_{0l} \exp (\boldsymbol{\gamma}_l^\top \bm{\omega}_{il}+\sum_{k=1}^K \alpha_{lk} \eta_{ik}(t|\bm{\beta}_k,\bm{b}_{ik})),
\end{eqnarray}
where $\bm{\omega}_{il}=(x_{1i},w_{1i},\ldots,w_{24,i})^\top$. Half of the covariates were independently generated from a standard normal distribution, while the others were independently generated from a binary distribution with a success probability of 0.5.
{To simulate time-to-event data, we apply the inverse of the cumulative hazard function \cite{austin2012generating} using the inverse probability integral transform method \cite{ross2012simulation}.}
\\
The covariance matrix of the random effects a structure is considered as follows:
\begin{eqnarray}\label{c}
\bm{\Sigma}=\begin{pmatrix}
\bm{\Sigma}^* & \bm{\Sigma}^\dagger  &  \ldots & \bm{\Sigma}^\dagger \\
  & \bm{\Sigma}^* & \ldots & \bm{\Sigma}^\dagger \\
  & & \ddots  & \vdots  \\
  & &  & \bm{\Sigma}^*
\end{pmatrix}, ~~ \rm{where}~~
\bm{\Sigma}^*=\begin{pmatrix}
1 & 0.5\\
0.5  & 1  
\end{pmatrix}, ~\bm{\Sigma}^\dagger=\rho\begin{pmatrix}
1 & 1\\
1 & 1  
\end{pmatrix}.
\end{eqnarray} 
For each scenario, $100$ data sets of $N=1000$ subjects were generated and split into 
a training sample of 500 subjects and a validation of 500 subjects.
The investigation focused on estimating parameters on the training sample using various approaches. The performance of these approaches for risk prediction was then evaluated using the validation samples. To achieve this goal, we consider four landmark times at $s=0, 0.25, 0.5, 0.75$, using $t=0.25$ as window for prediction. 
\subsection{Simulation with common effect for the two  events}
Among the 10 generated longitudinal markers and the 24 generated time-invariant covariates, only 4 markers and 4 covariates (2 Gaussian and 2 binary) were truly associated with the two event risks, sharing identical regression parameters for both events.  To evaluate the impact of increasing measurement errors, association parameters, and marker correlation on the performance of the three methods, we considered different values for these parameters, as described below:
\begin{enumerate}
   \item $\sigma_k^2=0.5,1$ for $k=1,2,\ldots,10$, 
   \item $\boldsymbol{\gamma}_l=(\gamma^*,-\gamma^*,\underbrace {0, \ldots ,0}_{10},\gamma^*,-\gamma^*,\underbrace {0, \ldots ,0}_{10})^\top,~l=1,2$ and $\bm{\alpha}_l=(-\alpha^*,-\alpha^*,\alpha^*,\alpha^*,0,0,0,0,0,0)^\top$, where $\gamma^*=\alpha^*=0.5,1$.
   \item $\rho=0.1,~0.7$.
\end{enumerate}
Note that the block-structured covariance matrix in~\eqref{c} is positive definite for all $\rho \in \{0.1, 0.7\}$.\\
Based on the insights obtained from both Table \ref{mcc}, which pertains to variable selection, and Table \ref{aucbs}, which focuses on risk prediction, regarding the performance of the VSJM, we can conclude:
    \begin{itemize}    

  \item For all scenarios, AUC and BS computed on the validation set after Bayesian variable selection on the training set were close to those of the oracle model.
    \item    Increasing the between markers correlation values tends to decrease the performance of the approach in terms of TPR, FPR, and MCC, albeit to an acceptable extent.

 \item Increasing the residual variance of the longitudinal markers ($\sigma^2$) leads to a decrease in the performance of the approach.

 \item Enhanced values of association parameters and regression coefficients tend to improve the approach's performance.

 \item 
Although the predictive performance of $BF$ and $lBFDR$ is similar, $BF$ generally outperforms $lBFDR$ in terms of total variable selection, except in the easiest scenario characterized by high association and low measurement error.


  \end{itemize}

\begin{table}
\footnotesize
 \centering
 \caption{Simulation results with identical effects for both events: Mean and standard deviation of the true positive rate (TPR), false positive rate (FPR), and Matthews Correlation Coefficient (MCC) over 100 replications. The results of DS and CS are based on a threshold of 1 for $BF$ and a threshold of 0.05 for $lBFDR$.
 \label{mcc}  }
\centering
\begin{tabular}{c|c|cc|cc|cc|cc|cc|cc}
  \hline
 & &\multicolumn{6}{c|}{CS } &\multicolumn{6}{c}{DS } \\ \hline
  & &\multicolumn{2}{c|}{Total}&\multicolumn{2}{c|}{L-Markers}&\multicolumn{2}{c|}{Fixed covariates } &\multicolumn{2}{c|}{Total}&\multicolumn{2}{c|}{L-Markers }&\multicolumn{2}{c}{Fixed covariates } \\ \hline
 &     & Est. & SD. & Est. & SD. & Est. & SD. & Est. & SD. & Est. & SD. & Est. & SD.   \\ 
 \hline
 \multicolumn{14}{c}{ $\rho=0.7,~\sigma^2=0.5,~\alpha^*=0.5,~\gamma^*=0.5$} \\
 \hline
$BF$  & TPR & 0.787 & 0.115 & 0.704 & 0.174 & 0.871 & 0.117 & 0.820 & 0.099 & 0.760 & 0.158 & 0.879 & 0.100 \\ 
& FPR & 0.028 & 0.023 & 0.049 & 0.060 & 0.021 & 0.025 & 0.053 & 0.029 & 0.097 & 0.088 & 0.040 & 0.032 \\ 
& MCC & 0.797 & 0.104 & 0.697 & 0.179 & 0.862 & 0.093 & 0.771 & 0.095 & 0.683 & 0.188 & 0.822 & 0.091 \\ 
   \hline
$lBFDR$ & TPR & 0.516 & 0.103 & 0.371 & 0.159 & 0.660 & 0.130 & 0.584 & 0.117 & 0.450 & 0.189 & 0.719 & 0.129 \\ 
& FPR & 0.001 & 0.005 & 0.001 & 0.011 & 0.001 & 0.005 & 0.005 & 0.010 & 0.012 & 0.030 & 0.003 & 0.008 \\ 
& MCC & 0.664 & 0.076 & 0.494 & 0.157 & 0.780 & 0.088 & 0.704 & 0.087 & 0.540 & 0.179 & 0.813 & 0.089 \\ 
\hline
  \multicolumn{14}{c}{$\rho=0.1,~\sigma^2=0.5,~\alpha^*=0.5,~\gamma^*=0.5$} \\
\hline
$BF$  & TPR & 0.941 & 0.054 & 1.000 & 0.000 & 0.881 & 0.109 & 0.947 & 0.054 & 1.000 & 0.000 & 0.894 & 0.108 \\ 
& FPR & 0.024 & 0.021 & 0.024 & 0.044 & 0.024 & 0.024 & 0.037 & 0.030 & 0.040 & 0.056 & 0.036 & 0.034 \\ 
& MCC & 0.913 & 0.051 & 0.973 & 0.050 & 0.860 & 0.090 & 0.892 & 0.069 & 0.954 & 0.063 & 0.840 & 0.109 \\ 
\hline
$lBFDR$ & TPR & 0.844 & 0.064 & 1.000 & 0.000 & 0.688 & 0.127 & 0.858 & 0.056 & 1.000 & 0.000 & 0.717 & 0.112 \\ 
& FPR & 0.001 & 0.004 & 0.001 & 0.011 & 0.001 & 0.005 & 0.003 & 0.007 & 0.006 & 0.021 & 0.002 & 0.008 \\ 
& MCC & 0.895 & 0.042 & 0.998 & 0.013 & 0.800 & 0.083 & 0.900 & 0.042 & 0.994 & 0.024 & 0.814 & 0.078 \\ 
\hline
  \multicolumn{14}{c}{$\rho=0.1,~\sigma^2=1,~\alpha^*=0.5,~\gamma^*=0.5$} \\
\hline
$BF$  & TPR & 0.929 & 0.058 & 1.000 & 0.000 & 0.858 & 0.116 & 0.939 & 0.057 & 1.000 & 0.000 & 0.877 & 0.114 \\ 
& FPR & 0.025 & 0.022 & 0.028 & 0.043 & 0.024 & 0.027 & 0.041 & 0.029 & 0.062 & 0.067 & 0.034 & 0.035 \\ 
& MCC & 0.903 & 0.054 & 0.967 & 0.050 & 0.848 & 0.095 & 0.880 & 0.066 & 0.930 & 0.073 & 0.834 & 0.113 \\ 
\hline
$lBFDR$ & TPR & 0.828 & 0.051 & 0.995 & 0.025 & 0.660 & 0.104 & 0.853 & 0.053 & 0.998 & 0.018 & 0.708 & 0.112 \\ 
& FPR & 0.003 & 0.009 & 0.005 & 0.020 & 0.003 & 0.011 & 0.002 & 0.006 & 0.005 & 0.020 & 0.001 & 0.005 \\ 
& MCC & 0.879 & 0.038 & 0.990 & 0.030 & 0.775 & 0.079 & 0.899 & 0.037 & 0.992 & 0.027 & 0.813 & 0.074 \\ 
\hline
\multicolumn{14}{c}{$\rho=0.1,~\sigma^2=0.5,~\alpha^*=1,~\gamma^*=1$} \\\hline
$BF$  & TPR & 0.999 & 0.009 & 1.000 & 0.000 & 0.998 & 0.018 & 0.999 & 0.009 & 1.000 & 0.000 & 0.998 & 0.018 \\ 
& FPR & 0.035 & 0.026 & 0.037 & 0.051 & 0.034 & 0.031 & 0.060 & 0.039 & 0.057 & 0.070 & 0.061 & 0.043 \\ 
& MCC & 0.933 & 0.048 & 0.958 & 0.058 & 0.914 & 0.075 & 0.890 & 0.064 & 0.936 & 0.077 & 0.856 & 0.088 \\ 
\hline
$lBFDR$ & TPR & 0.989 & 0.030 & 1.000 & 0.000 & 0.978 & 0.060 & 0.988 & 0.025 & 1.000 & 0.000 & 0.975 & 0.051 \\ 
& FPR & 0.003 & 0.008 & 0.002 & 0.012 & 0.003 & 0.008 & 0.007 & 0.012 & 0.008 & 0.025 & 0.007 & 0.013 \\ 
& MCC & 0.987 & 0.026 & 0.998 & 0.014 & 0.978 & 0.047 & 0.977 & 0.032 & 0.990 & 0.029 & 0.965 & 0.053 \\ 
\hline
\multicolumn{14}{c}{$\rho=0.1,~\sigma^2=1,~\alpha^*=1,~\gamma^*=1$} \\\hline
$BF$  & TPR & 0.996 & 0.016 & 1.000 & 0.000 & 0.992 & 0.032 & 1.000 & 0.000 & 1.000 & 0.000 & 1.000 & 0.000 \\ 
& FPR & 0.038 & 0.026 & 0.053 & 0.056 & 0.034 & 0.027 & 0.062 & 0.025 & 0.081 & 0.056 & 0.057 & 0.029 \\ 
& MCC & 0.924 & 0.045 & 0.939 & 0.063 & 0.908 & 0.064 & 0.885 & 0.041 & 0.908 & 0.060 & 0.862 & 0.062 \\ 
   \hline
$lBFDR$ & TPR & 0.979 & 0.038 & 1.000 & 0.000 & 0.958 & 0.076 & 0.981 & 0.029 & 1.000 & 0.000 & 0.963 & 0.058 \\ 
& FPR & 0.003 & 0.007 & 0.006 & 0.021 & 0.002 & 0.006 & 0.010 & 0.011 & 0.019 & 0.036 & 0.007 & 0.011 \\ 
& MCC & 0.981 & 0.028 & 0.994 & 0.025 & 0.970 & 0.051 & 0.968 & 0.027 & 0.977 & 0.042 & 0.958 & 0.046 \\ 
   \hline
\end{tabular}
\vspace*{30pt}
\end{table}

\begin{table}
\footnotesize
 \centering
 \caption{Simulation results with identical effects for both events: Mean and standard deviation of  AUC and Brier score computed for landmark times $s=0,0.25,0.5,0.75$ and prediction window of $0.25$. \label{aucbs}  }
\centering
\begin{tabular}{c|c|cc|cc|cccc|cccc}
  \hline
 & &\multicolumn{2}{c|}{Real } &\multicolumn{2}{c|}{Oracle } &\multicolumn{4}{c|}{CS } &\multicolumn{4}{c}{DS } \\ \hline
  & &\multicolumn{2}{c|}{ } &\multicolumn{2}{c|}{ } &\multicolumn{2}{c}{$BF$ } &\multicolumn{2}{c}{$lBFDR$ }&\multicolumn{2}{c}{$BF$ } &\multicolumn{2}{c}{$lBFDR$ } \\ \hline
 &   s & Est. & SD. & Est. & SD. & Est. & SD.  & Est. & SD. & Est. & SD. & Est. & SD. \\ 
 \hline
 \multicolumn{14}{c}{$\rho=0.7,~\sigma^2=0.5,~\alpha^*=0.5,~\gamma^*=0.5$} \\
\hline
AUC & 0 & 0.789 & 0.026 & 0.770 & 0.028 & 0.724 & 0.033 & 0.711 & 0.036 & 0.723 & 0.033 & 0.712 & 0.038 \\ 
& 0.25 & 0.737 & 0.051 & 0.719 & 0.048 & 0.685 & 0.052 & 0.665 & 0.051 & 0.685 & 0.051 & 0.674 & 0.051 \\ 
& 0.5 & 0.713 & 0.071 & 0.702 & 0.069 & 0.672 & 0.068 & 0.651 & 0.087 & 0.668 & 0.072 & 0.651 & 0.081 \\ 
& 0.75 & 0.702 & 0.126 & 0.681 & 0.123 & 0.653 & 0.123 & 0.635 & 0.137 & 0.645 & 0.122 & 0.632 & 0.138 \\ 
   \hline
BS & 0 & 0.160 & 0.011 & 0.161 & 0.010 & 0.170 & 0.012 & 0.171 & 0.013 & 0.170 & 0.013 & 0.171 & 0.014 \\ 
& 0.25 & 0.132 & 0.015 & 0.134 & 0.015 & 0.138 & 0.015 & 0.140 & 0.015 & 0.138 & 0.015 & 0.139 & 0.016 \\ 
& 0.5 & 0.114 & 0.027 & 0.114 & 0.027 & 0.117 & 0.027 & 0.119 & 0.027 & 0.117 & 0.027 & 0.119 & 0.027 \\ 
& 0.75 & 0.096 & 0.027 & 0.097 & 0.027 & 0.099 & 0.027 & 0.100 & 0.027 & 0.100 & 0.028 & 0.100 & 0.027 \\ 
   \hline
   \multicolumn{14}{c}{$\rho=0.1,~\sigma^2=0.5,~\alpha^*=0.5,~\gamma^*=0.5$} \\
  \hline
AUC & 0 & 0.850 & 0.021 & 0.816 & 0.025 & 0.803 & 0.024 & 0.796 & 0.027 & 0.801 & 0.026 & 0.795 & 0.028 \\ 
& 0.25 & 0.806 & 0.046 & 0.786 & 0.052 & 0.781 & 0.050 & 0.776 & 0.050 & 0.779 & 0.052 & 0.776 & 0.051 \\ 
& 0.5 & 0.796 & 0.064 & 0.778 & 0.067 & 0.770 & 0.063 & 0.768 & 0.066 & 0.773 & 0.065 & 0.771 & 0.065 \\ 
& 0.75  & 0.811 & 0.099 & 0.802 & 0.099 & 0.794 & 0.101 & 0.798 & 0.100 & 0.793 & 0.104 & 0.798 & 0.102 \\ 
\hline
BS & 0 & 0.154 & 0.008 & 0.159 & 0.009 & 0.161 & 0.009 & 0.162 & 0.010 & 0.161 & 0.009 & 0.162 & 0.010 \\ 
& 0.25 & 0.102 & 0.016 & 0.105 & 0.017 & 0.105 & 0.017 & 0.106 & 0.017 & 0.106 & 0.017 & 0.106 & 0.017 \\ 
& 0.5 & 0.077 & 0.018 & 0.078 & 0.018 & 0.079 & 0.018 & 0.078 & 0.018 & 0.079 & 0.018 & 0.078 & 0.018 \\ 
& 0.75  & 0.052 & 0.015 & 0.053 & 0.016 & 0.054 & 0.016 & 0.054 & 0.016 & 0.054 & 0.016 & 0.054 & 0.016 \\ 
\hline
   \multicolumn{14}{c}{$\rho=0.1,~\sigma^2=1,~\alpha^*=0.5,~\gamma^*=0.5$} \\
  \hline
AUC & 0 & 0.854 & 0.025 & 0.806 & 0.030 & 0.791 & 0.032 & 0.784 & 0.031 & 0.789 & 0.033 & 0.786 & 0.031 \\ 
& 0.25 & 0.812 & 0.047 & 0.790 & 0.054 & 0.781 & 0.054 & 0.776 & 0.053 & 0.783 & 0.057 & 0.780 & 0.052 \\ 
& 0.5 & 0.785 & 0.078 & 0.769 & 0.078 & 0.761 & 0.076 & 0.758 & 0.081 & 0.764 & 0.079 & 0.765 & 0.078 \\ 
& 0.75  & 0.811 & 0.108 & 0.791 & 0.104 & 0.779 & 0.109 & 0.771 & 0.110 & 0.777 & 0.116 & 0.779 & 0.109 \\ 
\hline
BS & 0 & 0.151 & 0.009 & 0.160 & 0.008 & 0.161 & 0.009 & 0.163 & 0.009 & 0.162 & 0.009 & 0.162 & 0.009 \\ 
& 0.25 & 0.107 & 0.014 & 0.110 & 0.015 & 0.110 & 0.015 & 0.110 & 0.015 & 0.110 & 0.015 & 0.110 & 0.015 \\ 
& 0.5 & 0.071 & 0.016 & 0.073 & 0.017 & 0.074 & 0.017 & 0.074 & 0.017 & 0.074 & 0.017 & 0.074 & 0.017 \\ 
& 0.75  & 0.054 & 0.015 & 0.055 & 0.015 & 0.056 & 0.015 & 0.056 & 0.015 & 0.055 & 0.015 & 0.056 & 0.015 \\ 
\hline
 \multicolumn{14}{c}{$\rho=0.1,~\sigma^2=0.5,~\alpha^*=1,~\gamma^*=1$} \\\hline
AUC & 0 & 0.942 & 0.014 & 0.891 & 0.022 & 0.883 & 0.018 & 0.885 & 0.019 & 0.881 & 0.019 & 0.885 & 0.020 \\ 
& 0.25 & 0.880 & 0.045 & 0.846 & 0.049 & 0.844 & 0.050 & 0.847 & 0.048 & 0.843 & 0.045 & 0.847 & 0.046 \\ 
& 0.5 & 0.890 & 0.050 & 0.860 & 0.066 & 0.861 & 0.077 & 0.863 & 0.073 & 0.863 & 0.068 & 0.866 & 0.063 \\ 
& 0.75 & 0.875 & 0.074 & 0.869 & 0.073 & 0.867 & 0.067 & 0.868 & 0.076 & 0.866 & 0.076 & 0.869 & 0.076 \\ 
   \hline
BS & 0 & 0.144 & 0.009 & 0.157 & 0.009 & 0.157 & 0.008 & 0.156 & 0.009 & 0.158 & 0.009 & 0.157 & 0.009 \\ 
& 0.25 & 0.064 & 0.011 & 0.067 & 0.013 & 0.068 & 0.012 & 0.068 & 0.012 & 0.068 & 0.011 & 0.068 & 0.012 \\ 
& 0.5 & 0.040 & 0.013 & 0.042 & 0.013 & 0.042 & 0.013 & 0.041 & 0.013 & 0.041 & 0.013 & 0.041 & 0.013 \\ 
& 0.75 & 0.028 & 0.012 & 0.028 & 0.012 & 0.028 & 0.011 & 0.028 & 0.011 & 0.028 & 0.011 & 0.028 & 0.011 \\ 
   \hline
 \multicolumn{14}{c}{$\rho=0.1,~\sigma^2=1,~\alpha^*=1,~\gamma^*=1$} \\\hline
AUC & 0 & 0.938 & 0.012 & 0.864 & 0.032 & 0.860 & 0.021 & 0.863 & 0.021 & 0.858 & 0.021 & 0.862 & 0.019 \\ 
& 0.25 & 0.889 & 0.030 & 0.843 & 0.041 & 0.847 & 0.033 & 0.849 & 0.032 & 0.842 & 0.031 & 0.849 & 0.030 \\ 
  & 0.5 & 0.858 & 0.057 & 0.826 & 0.069 & 0.828 & 0.054 & 0.832 & 0.054 & 0.829 & 0.048 & 0.835 & 0.053 \\ 
& 0.75 & 0.900 & 0.073 & 0.856 & 0.077 & 0.861 & 0.094 & 0.866 & 0.092 & 0.863 & 0.090 & 0.866 & 0.090 \\ 
   \hline
BS & 0 & 0.144 & 0.007 & 0.163 & 0.010 & 0.160 & 0.008 & 0.160 & 0.009 & 0.160 & 0.007 & 0.160 & 0.008 \\ 
& 0.25 & 0.063 & 0.012 & 0.069 & 0.013 & 0.068 & 0.012 & 0.067 & 0.012 & 0.068 & 0.012 & 0.067 & 0.012 \\ 
& 0.5 & 0.040 & 0.011 & 0.042 & 0.012 & 0.043 & 0.012 & 0.042 & 0.012 & 0.043 & 0.012 & 0.042 & 0.012 \\ 
& 0.75 & 0.028 & 0.010 & 0.030 & 0.011 & 0.030 & 0.010 & 0.030 & 0.010 & 0.030 & 0.010 & 0.030 & 0.010 \\ 
   \hline
\end{tabular}
\vspace*{30pt}
\end{table}
\FloatBarrier

\subsection{Scenarios with different  effects for two events}
In this section, we delve into exploring scenarios where the markers and covariates have differential effects on the  two causes of event. 
The simulation design was identical to section 4.1 except for the values of the regression parameters $\bm{\alpha}_l$ and $\boldsymbol{\gamma}_l$. Throughout all simulations within this context, we set $\sigma_k$ to 0.5 for $k$ ranging from 1 to 10, while $\gamma^*$ and $\alpha^*$ are fixed at 0.5, and $\rho$ is set to $0.1$. We consider three scenarios as follows:
\begin{description}
    \item[Single effects on cause 1] $~$\\
    \begin{itemize}
        \item $\bm{\alpha}_1=(-\alpha^*,-\alpha^*,\alpha^*,\alpha^*,0,0,0,0,0,0)^\top$ 
        \item $\boldsymbol{\gamma}_1=(\gamma^*,-\gamma^*,\underbrace {0, \ldots ,0}_{10},\gamma^*,-\gamma^*,\underbrace {0, \ldots ,0}_{10})^\top$. 
        \item $\bm{\alpha}_2=(0,...,0)^\top$
        \item $\boldsymbol{\gamma}_2=(\gamma^*,-\gamma^*,\underbrace {0, \ldots ,0}_{10},\gamma^*,-\gamma^*,\underbrace {0, \ldots ,0}_{10})^\top$. 
    \end{itemize}

    \item[Single effects on both causes] $~$\\
    \begin{itemize}
        \item $\bm{\alpha}_1=(\alpha^*,\alpha^*,0,0,0,0,0,0,0,0)^\top$ 
        \item $\boldsymbol{\gamma}_1=(\gamma^*,\gamma^*,\underbrace {0, \ldots ,0}_{22})^\top$. 
        \item $\bm{\alpha}_2=(0,0,\alpha^*,\alpha^*,0,0,0,0,0,0)^\top$ 
        \item $\boldsymbol{\gamma}_2=(\underbrace {0, \ldots ,0}_{12},\gamma^*,\gamma^*,\underbrace {0, \ldots ,0}_{10})^\top$. 
    \end{itemize}

    \item[Opposite effects on the two causes] $~$\\
    \begin{itemize}
        \item $\bm{\alpha}_1=(-\alpha^*,-\alpha^*,\alpha^*,\alpha^*,0,0,0,0,0,0)^\top$ 
        \item $\boldsymbol{\gamma}_1=(\gamma^*,-\gamma^*,\underbrace {0, \ldots ,0}_{10},\gamma^*,-\gamma^*,\underbrace {0, \ldots ,0}_{10})^\top$. 
        \item $\bm{\alpha}_2=(\alpha^*,\alpha^*,-\alpha^*,-\alpha^*,0,0,0,0,0,0)^\top$ 
        \item $\boldsymbol{\gamma}_2=(-\gamma^*,\gamma^*,\underbrace {0, \ldots ,0}_{10},-\gamma^*,\gamma^*,\underbrace {0, \ldots ,0}_{10})^\top$. 
    \end{itemize}
\end{description}
By analyzing the findings from Table \ref{mcc2}, which focuses on variable selection, and Table \ref{aucbs2}, which pertains to risk prediction, We observe that the differing association structures between the two events have minimal impact on the performance of the VSJM method.

\begin{table}
\footnotesize
 \centering
 \caption{Simulation results with differential effects between events: Mean and standard deviation of the true positive rate (TPR), false positive rate (FPR), and Matthews Correlation Coefficient (MCC) over 100 replications for $\rho=0.1,~\sigma^2=0.5,~\alpha^*=0.5,~\gamma^*=0.5$. The results of DS and CS are based on a threshold of 1 for $BF$ and a threshold of 0.05 for $lBFDR$.
 \label{mcc2}  }
\centering
\begin{tabular}{c|c|cc|cc|cc|cc|cc|cc}
  \hline
 & &\multicolumn{6}{c|}{CS } &\multicolumn{6}{c}{DS } \\ \hline
  & &\multicolumn{2}{c|}{Total}&\multicolumn{2}{c|}{L-Markers}&\multicolumn{2}{c|}{Fixed covariates } &\multicolumn{2}{c|}{Total}&\multicolumn{2}{c|}{L-Markers }&\multicolumn{2}{c}{Fixed covariates } \\ \hline
 &     & Est. & SD. & Est. & SD. & Est. & SD. & Est. & SD. & Est. & SD. & Est. & SD.   \\ 
 \hline
 \multicolumn{14}{c}{Single effects on cause 1} \\
 \hline
$BF$  & TPR  & 1.000 & 0.000 & 1.000 & 0.000 & 1.000 & 0.000 & 1.000 & 0.000 & 1.000 & 0.000 & 1.000 & 0.000 \\ 
& FPR & 0.018 & 0.018 & 0.022 & 0.049 & 0.017 & 0.020 & 0.039 & 0.024 & 0.056 & 0.060 & 0.035 & 0.023 \\ 
& MCC & 0.945 & 0.051 & 0.975 & 0.055 & 0.869 & 0.138 & 0.890 & 0.061 & 0.937 & 0.068 & 0.754 & 0.112 \\ 
   \hline
$lBFDR$ & TPR  & 1.000 & 0.000 & 1.000 & 0.000 & 1.000 & 0.000 & 1.000 & 0.000 & 1.000 & 0.000 & 1.000 & 0.000 \\ 
& FPR & 0.000 & 0.000 & 0.000 & 0.000 & 0.000 & 0.000 & 0.002 & 0.006 & 0.000 & 0.000 & 0.003 & 0.008 \\ 
& MCC & 1.000 & 0.000 & 1.000 & 0.000 & 1.000 & 0.000 & 0.993 & 0.019 & 1.000 & 0.000 & 0.974 & 0.068 \\ 
 \hline
  \multicolumn{14}{c}{Single effects on both causes} \\
\hline
$BF$  & TPR & 0.963 & 0.060 & 1.000 & 0.000 & 0.925 & 0.121 & 0.975 & 0.053 & 1.000 & 0.000 & 0.950 & 0.105 \\ 
& FPR & 0.020 & 0.027 & 0.050 & 0.087 & 0.009 & 0.016 & 0.035 & 0.032 & 0.069 & 0.080 & 0.023 & 0.024 \\ 
& MCC & 0.908 & 0.123 & 0.914 & 0.144 & 0.912 & 0.116 & 0.868 & 0.109 & 0.874 & 0.132 & 0.867 & 0.130 \\ 
\hline
$lBFDR$ & TPR & 0.863 & 0.092 & 1.000 & 0.000 & 0.725 & 0.184 & 0.887 & 0.071 & 1.000 & 0.000 & 0.775 & 0.142 \\ 
& FPR & 0.000 & 0.000 & 0.000 & 0.000 & 0.000 & 0.000 & 0.003 & 0.011 & 0.013 & 0.040 & 0.000 & 0.000 \\ 
& MCC & 0.919 & 0.055 & 1.000 & 0.000 & 0.836 & 0.115 & 0.922 & 0.060 & 0.976 & 0.075 & 0.869 & 0.086 \\ 
\hline
  \multicolumn{14}{c}{Opposite effects  on the two causes} \\
\hline
$BF$  & TPR & 0.981 & 0.030 & 1.000 & 0.000 & 0.963 & 0.060 & 0.988 & 0.026 & 1.000 & 0.000 & 0.975 & 0.053 \\ 
& FPR  & 0.015 & 0.015 & 0.025 & 0.040 & 0.013 & 0.018 & 0.044 & 0.027 & 0.058 & 0.056 & 0.040 & 0.038 \\ 
& MCC & 0.956 & 0.044 & 0.971 & 0.047 & 0.943 & 0.066 & 0.908 & 0.049 & 0.933 & 0.063 & 0.887 & 0.093  \\ 
\hline
$lBFDR$ & TPR & 0.850 & 0.053 & 1.000 & 0.000 & 0.700 & 0.105 & 0.881 & 0.069 & 1.000 & 0.000 & 0.762 & 0.138 \\ 
& FPR  & 0.000 & 0.000 & 0.000 & 0.000 & 0.000 & 0.000 & 0.006 & 0.013 & 0.000 & 0.000 & 0.008 & 0.017 \\ 
& MCC & 0.901 & 0.035 & 1.000 & 0.000 & 0.811 & 0.070 & 0.910 & 0.055 & 1.000 & 0.000 & 0.829 & 0.105  \\ 
\hline
\end{tabular}
\vspace*{30pt}
\end{table}

\begin{table}
\footnotesize
 \centering
 \caption{Simulation results with differential effects between events: Mean and standard deviation of  AUC and Brier score computed for landmark times $s=0,0.25,0.5,0.75$ and prediction window of $0.25$ for $\rho=0.1,~\sigma^2=0.5,~\alpha^*=0.5,~\gamma^*=0.5$. \label{aucbs2}  }
\centering
\begin{tabular}{c|c|cc|cc|cccc|cccc}
  \hline
 & &\multicolumn{2}{c|}{Real } &\multicolumn{2}{c|}{Oracle } &\multicolumn{4}{c|}{CS } &\multicolumn{4}{c}{DS } \\ \hline
  & &\multicolumn{2}{c|}{ } &\multicolumn{2}{c|}{ } &\multicolumn{2}{c}{$BF$ } &\multicolumn{2}{c}{$lBFDR$ }&\multicolumn{2}{c}{$BF$ } &\multicolumn{2}{c}{$lBFDR$ } \\ \hline
 &   s & Est. & SD. & Est. & SD. & Est. & SD.  & Est. & SD. & Est. & SD. & Est. & SD. \\ 
 \hline
 \multicolumn{14}{c}{Single effects on cause 1} \\
\hline
AUC & 0 &   0.781 & 0.029 & 0.747 & 0.026 & 0.742 & 0.019 & 0.740 & 0.020 & 0.742 & 0.017 & 0.737 & 0.020 \\ 
 &   0.25 & 0.790 & 0.047 & 0.759 & 0.060 & 0.761 & 0.050 & 0.762 & 0.053 & 0.757 & 0.052 & 0.757 & 0.053 \\ 
 &   0.5 & 0.786 & 0.063 & 0.766 & 0.057 & 0.769 & 0.054 & 0.772 & 0.058 & 0.769 & 0.059 & 0.771 & 0.066 \\ 
 &   0.75 & 0.778 & 0.090 & 0.794 & 0.087 & 0.786 & 0.095 & 0.796 & 0.090 & 0.787 & 0.089 & 0.790 & 0.092 \\    \hline
BS & 0  & 0.147 & 0.011 & 0.153 & 0.010 & 0.157 & 0.008 & 0.153 & 0.010 & 0.157 & 0.008 & 0.154 & 0.009 \\ 
 &   0.25 & 0.117 & 0.014 & 0.121 & 0.017 & 0.121 & 0.016 & 0.120 & 0.017 & 0.121 & 0.016 & 0.121 & 0.017 \\ 
 &   0.5 & 0.075 & 0.019 & 0.077 & 0.019 & 0.077 & 0.017 & 0.077 & 0.019 & 0.077 & 0.018 & 0.076 & 0.019 \\ 
 &   0.75 & 0.063 & 0.017 & 0.063 & 0.018 & 0.063 & 0.017 & 0.063 & 0.018 & 0.063 & 0.018 & 0.064 & 0.018 \\ \hline
   \multicolumn{14}{c}{Single effects on both causes} \\
  \hline
AUC & 0 & 0.733 & 0.027 & 0.729 & 0.026 & 0.723 & 0.021 & 0.708 & 0.026 & 0.723 & 0.020 & 0.709 & 0.025 \\ 
 &   0.25 & 0.762 & 0.035 & 0.752 & 0.036 & 0.743 & 0.035 & 0.734 & 0.038 & 0.746 & 0.034 & 0.736 & 0.037 \\ 
 &   0.5 & 0.756 & 0.073 & 0.743 & 0.082 & 0.736 & 0.079 & 0.725 & 0.071 & 0.733 & 0.079 & 0.736 & 0.077 \\ 
 &   0.75 & 0.756 & 0.073 & 0.730 & 0.071 & 0.741 & 0.064 & 0.736 & 0.056 & 0.730 & 0.066 & 0.743 & 0.060 \\ 
\hline
BS & 0 & 0.149 & 0.011 & 0.151 & 0.011 & 0.156 & 0.011 & 0.156 & 0.010 & 0.156 & 0.011 & 0.156 & 0.010 \\ 
 &   0.25 & 0.134 & 0.017 & 0.136 & 0.017 & 0.138 & 0.016 & 0.139 & 0.017 & 0.137 & 0.017 & 0.139 & 0.017 \\ 
 &   0.5 & 0.124 & 0.017 & 0.124 & 0.018 & 0.127 & 0.018 & 0.128 & 0.017 & 0.128 & 0.018 & 0.126 & 0.017 \\ 
 &   0.75 & 0.100 & 0.020 & 0.106 & 0.020 & 0.104 & 0.018 & 0.106 & 0.019 & 0.106 & 0.019 & 0.105 & 0.019 \\ 
\hline
   \multicolumn{14}{c}{Opposite effects  on the two causes} \\
  \hline
AUC & 0 & 0.784 & 0.021 & 0.791 & 0.027 & 0.765 & 0.027 & 0.757 & 0.026 & 0.761 & 0.029 & 0.758 & 0.030 \\ 
&   0.25 & 0.762 & 0.042 & 0.762 & 0.039 & 0.753 & 0.048 & 0.745 & 0.048 & 0.745 & 0.039 & 0.745 & 0.040 \\ 
 &   0.5 & 0.758 & 0.078 & 0.735 & 0.073 & 0.716 & 0.070 & 0.710 & 0.065 & 0.720 & 0.056 & 0.697 & 0.060 \\ 
 &   0.75 & 0.732 & 0.112 & 0.691 & 0.105 & 0.710 & 0.115 & 0.693 & 0.137 & 0.706 & 0.114 & 0.706 & 0.132 \\  
\hline
BS & 0 & 0.138 & 0.011 & 0.137 & 0.011 & 0.146 & 0.012 & 0.149 & 0.011 & 0.147 & 0.012 & 0.149 & 0.012 \\ 
&   0.25 & 0.136 & 0.017 & 0.137 & 0.019 & 0.140 & 0.021 & 0.143 & 0.022 & 0.143 & 0.020 & 0.143 & 0.022 \\ 
 &   0.5 & 0.145 & 0.029 & 0.153 & 0.028 & 0.156 & 0.026 & 0.160 & 0.025 & 0.159 & 0.023 & 0.164 & 0.025 \\ 
 &   0.75 & 0.145 & 0.034 & 0.150 & 0.038 & 0.148 & 0.034 & 0.154 & 0.032 & 0.158 & 0.039 & 0.157 & 0.033 \\ 
\hline
\end{tabular}
\vspace*{30pt}
\end{table}
\FloatBarrier

\section{Application: 3C study}
\subsection{Data description}
\begin{figure}[ht]
\centering
\includegraphics[width=12cm]{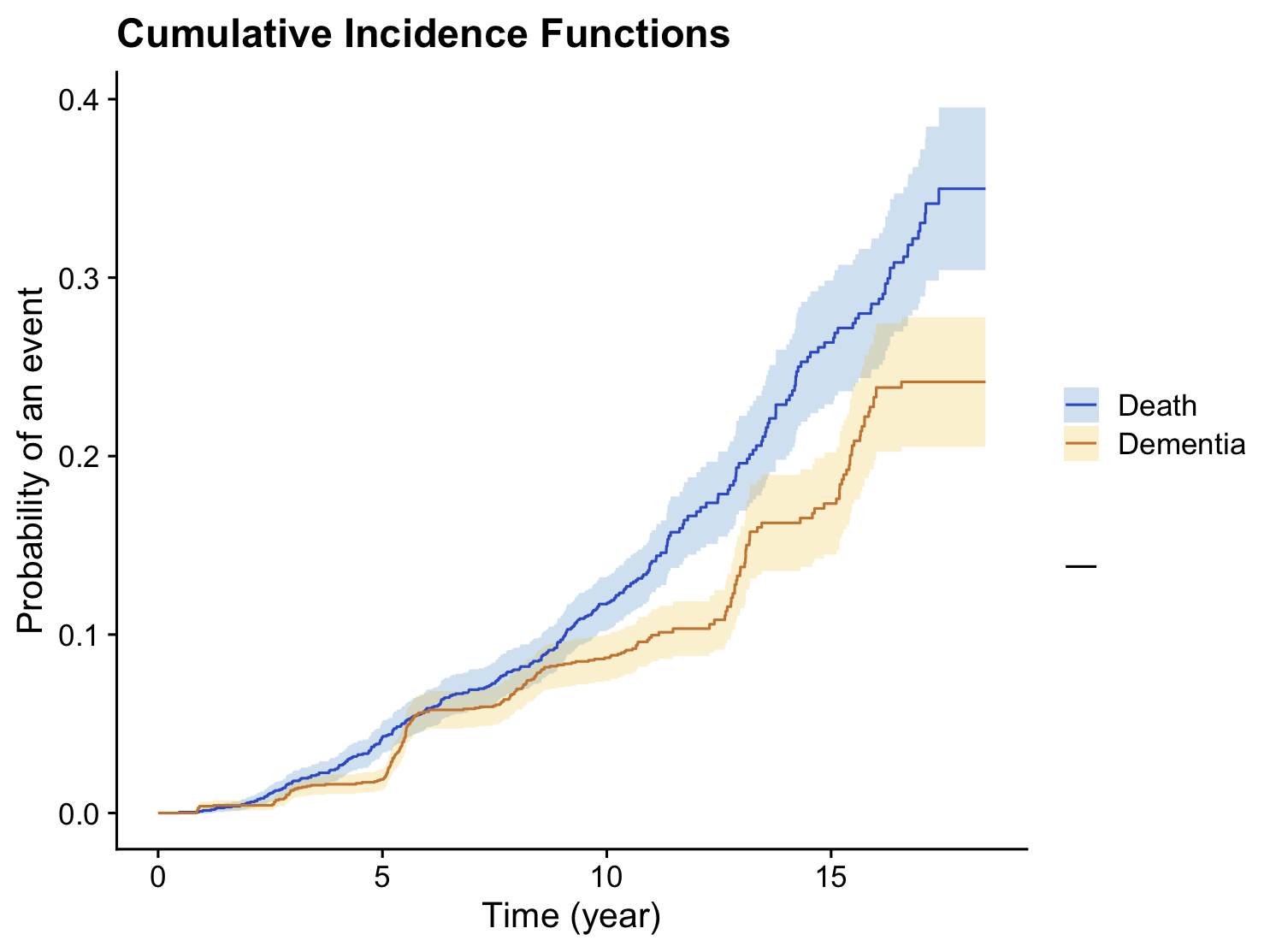}
\vspace*{-.1cm} \caption{\label{cif} Cumulative incidence function for dementia and death event over 17 years of follow-up. }
\end{figure}

\begin{landscape}
\begin{figure}[ht]
\centering
\includegraphics[width=21cm]{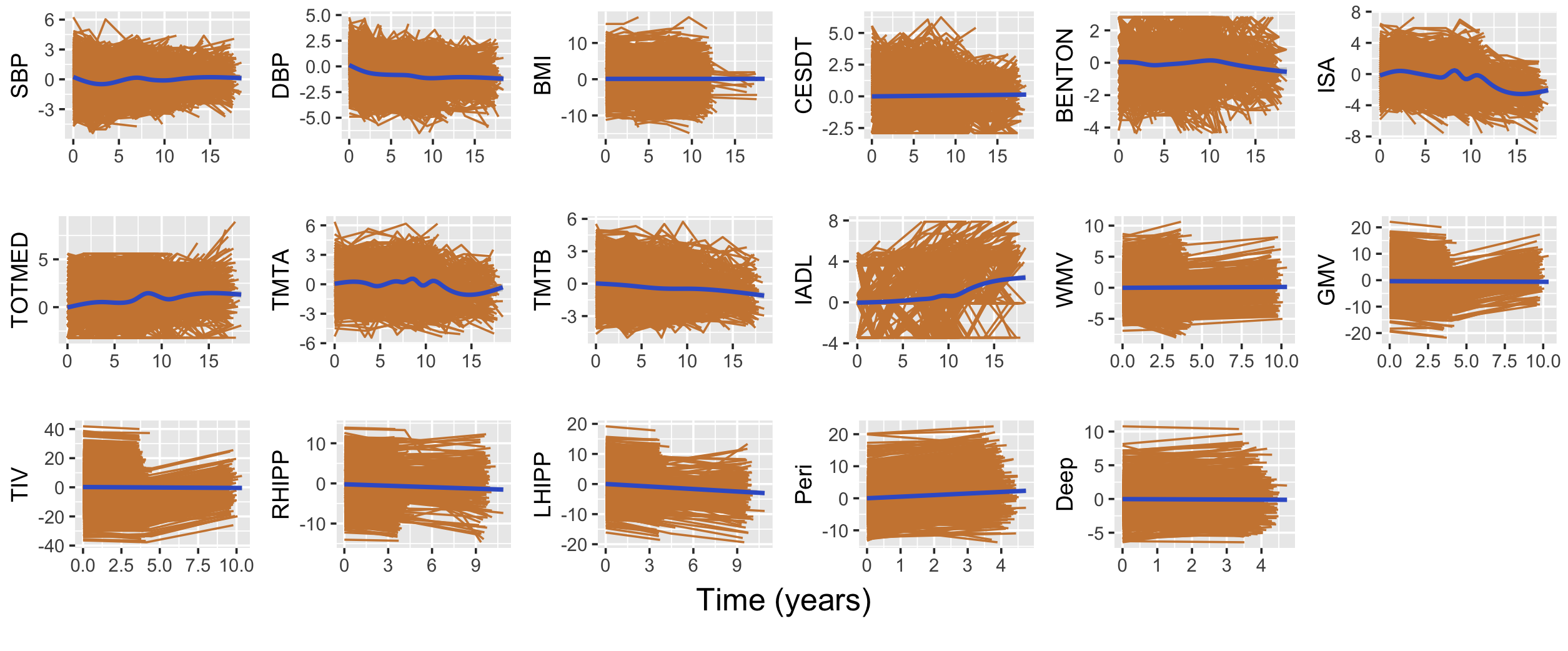}
\vspace*{-.1cm} \caption{\label{sp} Individual trajectories from the normalized 
clinical, neuropsychological and imaging longitudinal markers in the 3C study. Systolic blood pressure (SBP), 
diastolic  blood pressure (DBP), body mass index (BMI),  depressive symptomatology  measured  
using the Center for Epidemiologic Studies Depression scale  (CESDT), the visual retention test of Benton (BENTON), Isaacs Set Test (ISA), the total number of medications (TOTMED), the trail making test A and B (TMTA and TMTB), functional dependency assessed using Instrumental Activity of Daily Living scale (IADL),
white matter volume (WMV), gray matter volume (GMV),  total intracranial volume  (TIV),
 right and left hippocampal volume (RHIPP and LHIPP), 
for the 5-year risk of dementia of thevolumes of White Matter Hyperintensities in the periventricular (Peri) and deep (Deep) white matter.}
\end{figure}
\end{landscape}

In this section, we analyzed a subset of the Three-City (3C) Study \citep{antoniak2003vascular} which is a French cohort study. Participants aged 65 or older were recruited from three cities (Bordeaux, Dijon, and Montpellier) and followed for over 10 years to investigate the association between vascular factors and dementia. The data includes socio-demographic information, general health information, and cognitive test scores. The application here only included the centers of Dijon and Bordeaux, where MRI exams were conducted at years 0 and 4 for both, and additionally at year 10 for Bordeaux. Except for the MRI markers, the other longitudinal markers were measured at baseline and over the follow-up visit, including years 2, 4, 7, 10, 12, and also 14 and 17 in Bordeaux only.
The diagnosis of dementia among participants followed a three-step process. Initially, cognitive assessments were conducted by neuropsychologists or investigators. Suspected cases of dementia then underwent examination by clinicians, with validation by an independent expert committee comprising neurologists and geriatricians. In addition to dementia, the exact time of death was collected and is considered a competing risk.\\
We considered $N = 2133$ subjects who were dementia-free at the beginning of the study and had at least one measurement for each of the longitudinal markers. Out of these subjects, 227 were diagnosed with incident dementia, and 311 died before      
developing dementia. Figure \ref{cif} shows the cumulative incidence function for dementia and death in this sample.\\
We considered seventeen longitudinal markers: three cardio-metabolic markers (body mass index (BMI), diastolic blood pressure (DBP), and systolic blood pressure (SBP)), the total number of medications (TOTMED), depressive symptomatology measured using the Center for Epidemiologic Studies-Depression scale (CESDT, the lower the better), functional dependency assessed using the Instrumental Activities of Daily Living scale (IADL, the lower the better), four cognitive tests (the visual retention test of Benton (BENTON, number of correct responses out of 15), the Isaacs set test (ISA, total number of words given in 4 semantic categories in 15 seconds), the trail making tests A and B (TMTA and TMTB, number of correct moves per minute), the total intracranial volume (TIV), and some biomarkers of neurodegeneration, including white matter volume (WMV), gray matter volume (GMV), left hippocampal volume (LHIPP), and right hippocampal volume (RHIPP); two markers of vascular brain lesions including the volumes of white matter hyperintensities in the periventricular (Peri), and deep (Deep) white matter.
Figure \ref{sp} shows individual trajectories of the longitudinal markers in the 3C study, illustrating time-dependent variables. The data were pre-transformed using splines to satisfy the normality assumption of linear mixed models
\citep{devaux2022random}.
The minimum, maximum, and median number of repeated measurements for SBP, DBP, CESDT, BENTON, ISA, TOTMED, and IADL are 1, 8, and 5, respectively. For BMI, TMTA, and TMTB, the minimum, maximum, and median number of repeated measurements are 1, 7, and 4, respectively. The minimum, maximum, and median number of repeated measurements for WMV, GMV, TIV, RHIPP, and LHIPP are 1, 3, and 2, respectively. For Peri and Deep (which were not measured at year 10), there are at most two repeated measures, with a median equal to 2.\\
In addition, the model incorporates five time-invariant explanatory variables including age at enrollment ($\frac{\rm{age-65}}{10}$), sex (0=male, 1=female), education level (Educ, 0=less than 10 years of schooling, 1=at least 10 years of schooling), Diabetes status at baseline (1=yes), and APOE4 allele carrier status (the main genetic susceptibility factor for Alzheimer's disease, 1=presence of APOE4). 

\FloatBarrier
\subsection{Data analysis}
For analyzing the 3C data using the proposed approach, we partitioned the dataset into two distinct sets: the training set, comprising 70\% of the data, and a separate validation set. Parameter estimation was conducted using the training data, while risk prediction was exclusively performed on the validation set.\\
 For this aim, a quadratic time trend is considered for both fixed and random effects, as a non-linear time trend was previously observed for SBP, DBP, BMI, CESDT, BENTON, ISA, TOTMED, TMTA, TMTB, and IADL \citep{devaux2022random}:
\begin{eqnarray}\label{13}
Y_{ijk}&=&\eta_{ik}(s_{ijk}|\bm{\beta}_k,\bm{b}_{ik})+\varepsilon_{ijk}\\\nonumber
&=&\beta_{0k}+\beta_{1k}s_{ijk}+\beta_{2k}s_{ijk}^2+
b_{0ik}+b_{1ik} s_{ijk}+b_{1ik} s_{ijk}^2+\varepsilon_{ijk}.\nonumber
\end{eqnarray}
For other longitudinal markers, a linear mixed model is considered:
\begin{eqnarray}\label{12}
Y_{ik}(t)&=&\eta_{ik}(s_{ijk}|\bm{\beta}_k,\bm{b}_{ik})+\varepsilon_{ijk}\\\nonumber
&=&\beta_{0k}+\beta_{1k}s_{ijk}+
b_{0ik}+b_{1ik} s_{ijk}+\varepsilon_{ijk}.\nonumber
\end{eqnarray}
To model dementia and death, we utilize proportional hazard models that may depend on the five time-invariant explanatory variables and the current value of the markers. Let $\lambda_{il}(t)$ denote the cause-specific hazard function for subject $i$, where $l=1$ for dementia and $l=2$ for death. The proportional hazard model takes the following form:
\begin{eqnarray}\label{hh}
\lambda_{il}(t)&=&\lambda_{0l}(t)
\exp(\gamma_{1l}APOE_{i}+\gamma_{2l}Sex_{i}+\gamma_{3l}Diabete_{i}+ \gamma_{4l}Educ_{i}\\\nonumber &+&\gamma_{5l}Age_{i}
+\sum_{k=1}^{18} \alpha_{kl} \eta_{ik}(t|\bm{\beta}_k,\bm{b}_{ik})).\nonumber
\end{eqnarray}
Table \ref{c1} in the supplementary material displays the estimated regression coefficients and variance of error from the mixed effects sub-models for each marker in the one-marker joint models. Additionally, Table \ref{c2} presents the estimated parameters from the proportional hazard models for dementia and death, considering the VSJM approach given you choose DS, this is an additional reason for not emphasizing on  a vague  superiority of CS which is debatable in the simulation results, including estimates, standard
deviations, 95\% credible intervals, $lBFDR$s, and $BF$s.
The analysis identifies several significant factors associated with dementia and death. 
Specifically, high TIV and IADL values, coupled with lower scores on the BENTON, ISA, and TMTB tests, are linked to an increased risk of dementia. Additionally, higher educational attainment, when adjusted for current cognitive scores, is also associated with a higher risk of dementia.
For death without dementia, lower DBP and BENTON scores, increasing age, diabetes, and high educational levels (adjusted for current cognitive scores) are associated with a higher risk. Conversely, being female are linked to a lower risk of death. The results for CS remains   similar, and are reported in Table \ref{c21}.\\
The predictive abilities of the model for the $t=5$ years prediction of dementia accounting for the competing risk of death were assessed at landmark times $s=0,5$ and $10$ years. To avoid overoptimistic results, AUC and BS were computed on a validation set. 
The predictive abilities presented in table  \ref{auc3c} are very good. The AUC is always higher than  0.9  except for s=0 with CS and $BF$ and higher for $s= 5$ and $10$ compared to $s=0$, highlighting the information brought by the repeated measures of the longitudinal  markers. Results were very similar for CS and DS priors and slightly better for $lBFDR$ compared to $BF$.

 \begin{table}[ht] 
 \caption{\label{c2} The estimated regression coefficients and variance of errors of the longitudinal sub-model from the joint model in 3C study data using DS. Est: posterior mean, SD: standard deviation, 2.5$\% $ CI: lower bound of credible interval  and  97.5$\% $ CI: upper  bound of credible interval.}
\centering
\scriptsize
\begin{tabular}{ccccccc}
  \hline
 & Est. & SD & $2.5\%$ CI & $97.5\%$ CI & $lBFDR$ & $BF$  \\ \hline
 &\multicolumn{6}{c}{Dementia} \\ 
 WMV & 0.000 & 0.000 & 0.000 & 0.000 & 1.000 & 0.000 \\ 
 GMV & 0.000 & 0.000 & 0.000 & 0.000 & 1.000 & 0.000 \\ 
 TIV & 0.060 & 0.010 & 0.040 & 0.079 & 0.000 & Inf \\ 
 RHIPP & 0.000 & 0.000 & 0.000 & 0.000 & 1.000 & 0.000 \\ 
 LHIPP & -0.079 & 0.018 & -0.115 & -0.043 & 0.000 & Inf \\ 
 Peri & 0.000 & 0.000 & 0.000 & 0.000 & 1.000 & 0.000 \\ 
 Deep & 0.000 & 0.000 & 0.000 & 0.000 & 1.000 & 0.000 \\ 
 SBP & 0.000 & 0.000 & 0.000 & 0.000 & 1.000 & 0.000 \\ 
 DBP & 0.000 & 0.000 & 0.000 & 0.000 & 1.000 & 0.000 \\ 
 BMI & -0.018 & 0.034 & -0.103 & 0.000 & 0.765 & 0.308 \\ 
 CESDT & 0.013 & 0.043 & 0.000 & 0.171 & 0.900 & 0.111 \\ 
 BENTON & -0.500 & 0.119 & -0.744 & -0.272 & 0.000 & Inf \\ 
 ISA & -0.323 & 0.079 & -0.469 & -0.171 & 0.000 & Inf \\ 
 TOTMED & 0.000 & 0.000 & 0.000 & 0.000 & 1.000 & 0.000 \\ 
 TMTA & 0.000 & 0.000 & 0.000 & 0.000 & 1.000 & 0.000 \\ 
 TMTB & -1.140 & 0.110 & -1.345 & -0.930 & 0.000 & Inf \\ 
 IADL & 0.326 & 0.062 & 0.207 & 0.445 & 0.000 & Inf \\ 
 Sex & 0.002 & 0.028 & 0.000 & 0.000 & 0.986 & 0.014 \\ 
 Diabete & 0.000 & 0.025 & 0.000 & 0.000 & 0.996 & 0.004 \\ 
 APOE & 0.002 & 0.032 & 0.000 & 0.000 & 0.994 & 0.006 \\ 
 Educ & 0.891 & 0.186 & 0.536 & 1.245 & 0.000 & Inf \\ 
 Age & 0.000 & 0.000 & 0.000 & 0.000 & 1.000 & 0.000 \\ 
 &\multicolumn{6}{c}{Death} \\ 
 WMV & 0.000 & 0.000 & 0.000 & 0.000 & 1.000 & 0.000 \\ 
 GMV & 0.000 & 0.000 & 0.000 & 0.000 & 1.000 & 0.000 \\ 
 TIV & 0.000 & 0.000 & 0.000 & 0.000 & 1.000 & 0.000 \\ 
 RHIPP & 0.000 & 0.000 & 0.000 & 0.000 & 1.000 & 0.000 \\ 
 LHIPP & 0.000 & 0.000 & 0.000 & 0.000 & 1.000 & 0.000 \\ 
 Peri & 0.000 & 0.000 & 0.000 & 0.000 & 1.000 & 0.000 \\ 
 Deep & 0.000 & 0.000 & 0.000 & 0.000 & 1.000 & 0.000 \\ 
 SBP & 0.000 & 0.000 & 0.000 & 0.000 & 1.000 & 0.000 \\ 
 DBP & -0.155 & 0.048 & -0.248 & -0.060 & 0.000 & Inf \\ 
 BMI & 0.000 & 0.000 & 0.000 & 0.000 & 1.000 & 0.000 \\ 
 CESDT & 0.000 & 0.000 & 0.000 & 0.000 & 1.000 & 0.000 \\ 
 BENTON & -3.347 & 0.132 & -3.623 & -3.097 & 0.000 & Inf \\ 
 ISA & 0.000 & 0.000 & 0.000 & 0.000 & 1.000 & 0.000 \\ 
 TOTMED & 0.000 & 0.000 & 0.000 & 0.000 & 1.000 & 0.000 \\ 
 TMTA & 0.000 & 0.000 & 0.000 & 0.000 & 1.000 & 0.000 \\ 
 TMTB & 0.000 & 0.000 & 0.000 & 0.000 & 1.000 & 0.000 \\ 
 IADL & 0.000 & 0.000 & 0.000 & 0.000 & 1.000 & 0.000 \\ 
 Sex & -2.200 & 0.175 & -2.553 & -1.872 & 0.000 & Inf \\ 
 Diabete & 1.046 & 0.183 & 0.677 & 1.398 & 0.000 & Inf \\  
 APOE & 0.000 & 0.000 & 0.000 & 0.000 & 1.000 & 0.000 \\ 
 Educ & 0.870 & 0.159 & 0.548 & 1.178 & 0.000 & Inf \\ 
 Age & 1.230 & 0.191 & 0.844 & 1.588 & 0.000 & Inf \\ 
\hline
\end{tabular}
\end{table}
 \FloatBarrier

\begin{table}[ht] 
 \caption{\label{auc3c} AUC and Brier score computed by validation set  for landmark times $s=0,5,10$ years and prediction window of 5 years for the 3C data.  }
\centering
\footnotesize 
\begin{tabular}{cccccccccc}
  \hline
   & &\multicolumn{4}{c}{$lBFDR$ }&\multicolumn{4}{c}{BF}  \\\hline
  & &\multicolumn{2}{c}{AUC }&\multicolumn{2}{c}{BS}  &\multicolumn{2}{c}{AUC }&\multicolumn{2}{c}{BS}  \\ \hline
 & $s$ & Est. & SD. & Est. & SD. & Est. & SD. & Est. & SD.  \\ 
   \hline
DS    &  \\ 
 & $0$ &  0.891 &  0.037 &  0.019 & 0.005  &  0.885 &  0.034 &  0.019 &   0.005    \\ 
 & $5$ &  0.891 &  0.018 &  0.130 &  0.014  &  0.846 & 0.027 & 0.152 & 0.016  \\
 & $10$ & 0.971 &  0.016 &   0.044 &  0.016  & 0.901  &  0.035  &  0.122  & 0.029 \\ 
  CS    &  \\ 
   & $0$ & 0.731 &  0.070 & 0.058 &  0.006 &  0.827 &  0.075 &  0.020 & 0.004 \\ 
 & $5$ & 0.851 &  0.022 &   0.184 & 0.016 &  0.895 & 0.017 &  0.109 &  0.012\\ 
 & $10$ & 0.863 &  0.045 &   0.135 &  0.028  &  0.945 &  0.023 &  0.086 &  0.022 \\ 
  \hline 
\end{tabular}
\end{table}

\section{Conclusion}
In this paper, we introduce a novel Bayesian approach for variable selection in joint modeling of multiple longitudinal markers and competing risks outcomes. Our method employs spike-and-slab priors for both the association parameters and the regression coefficients in the survival model. The approach involves two main steps: first, a simple Bayesian joint model is fitted separately for each longitudinal marker and competing risks outcome; second, variable selection is performed using a time-dependent Bayesian proportional hazards model, incorporating Dirac spike or continuous spike priors for the regression coefficients, including those for time-dependent markers and fixed covariates. We also propose two decision criteria, Bayes factor and local Bayesian false discovery rate ($lBFDR$), for selecting relevant variables.  
\\
A key advantage of our Bayesian variable selection approach is that the posterior distributions of the selected variables naturally account for both the uncertainty of selection and parameter estimation, without requiring re-estimation of the model using only the selected variables. This property is in contrast to frequentist post-selection approaches, where ignoring the selection uncertainty can lead to underestimated standard errors \cite{George1993,O'Hara2009,Malsiner-Walli2018}. Consequently, the standard deviations obtained from our approach provide valid measures of uncertainty for both selection and estimation, even in the presence of multiple imputed datasets or informative dropout.  
\\
In addition to variable selection, we illustrate how our approach can be used for dynamic prediction based on the estimated joint model. We demonstrate the performance of the method via extensive simulation studies, evaluating both variable selection accuracy and predictive ability. Furthermore, we apply our method to predict dementia risk using the 3C dataset, showcasing its practical relevance. The method is implemented in the freely available R package \texttt{VSJM} (\url{https://github.com/tbaghfalaki/VSJM}).  
\\
Future extensions of this work could include accommodating non-Gaussian longitudinal markers by using generalized linear mixed models in the one-marker joint models. Additionally, the methodology could be expanded to perform group-wise selection, providing further flexibility for high-dimensional longitudinal data. Overall, our approach offers a computationally efficient and statistically principled framework for variable selection and dynamic prediction in joint modeling of longitudinal and time-to-event data.






 \subsubsection*{Acknowledgements}
This work was partly funded by the French National Research Agency (grant ANR-21-CE36 for the project “Joint Models for
Epidemiology and Clinical research”). This study received financial support from the French government in the framework of the University of Bordeaux's IdEx "Investments for the Future" program / RRI PHDS.
The 3C Study was supported by Sanofi-Synthélabo, the FRM, the CNAM-TS,DGS,Conseils Régionaux of Aquitaine, 395
Languedoc-Roussillon, and Bourgogne; Foundation of France; Ministry of Research- INSERM “Cohorts and biological data
collections" program; MGEN; Longevity Institute; General Council of the Côte d’Or; ANR PNRA 2006 (grant ANR/ DEDD/
PNRA/ PROJ/ 200206–01-01) and Longvie 2007 (grant LVIE-003-01); Alzheimer Plan Foundation (FCS grant 2009-2012);
and Roche. The Three City Study data are available upon request at \url{e3c.coordinatingcenter@gmail.com}.

 \bibliography{sn-bibliography}

\clearpage
\newpage
\section*{Supplementary Material}
\setcounter{table}{0}
\renewcommand{\thetable}{A.\arabic{table}}
\setcounter{figure}{0}
\renewcommand{\thefigure}{A.\arabic{figure}}

\begin{table}[ht] 
\centering
 \caption{\label{c1} Estimated regression coefficients and error variances from the longitudinal sub-model in the 3C study. Est.: posterior mean; SD: standard deviation; 2.5\% CI: lower bound of credible bound; 97.5\% CI: upper bound of credible bound.}
\footnotesize
        \centering
\begin{tabular}{c|c|cccc}
  \hline
& & Est. & SD & $2.5\%$ CI & $97.5\%$ CI  \\   \hline
WMV & \rm{Intercept} & -0.004 & 0.063 & -0.119 & 0.114 \\ 
  & \rm{Time} & 0.027 & 0.012 & 0.004 & 0.051 \\ 
   & $\sigma^2$ & 0.765 & 0.063 & 0.629 & 0.875 \\  \hline
GMV  & \rm{Intercept} & -0.102 & 0.145 & -0.382 & 0.170 \\ 
 &  \rm{Time} & -0.246 & 0.021 & -0.285 & -0.205 \\ 
 &  $\sigma^2$ & 1.994 & 0.107 & 1.797 & 2.216 \\  \hline
TIV & \rm{Intercept} & -0.293 & 0.299 & -0.779 & 0.420 \\ 
 &  \rm{Time} & 0.019 & 0.028 & -0.037 & 0.074 \\ 
 &  $\sigma^2$ & 3.760 & 0.199 & 3.394 & 4.168 \\  \hline
RHIPP & \rm{Intercept} & 0.137 & 0.100 & -0.042 & 0.327 \\ 
 & \rm{Time} & -0.316 & 0.013 & -0.340 & -0.291 \\ 
 &  $\sigma^2$ & 0.905 & 0.054 & 0.801 & 1.018 \\  \hline
LHIPP & \rm{Intercept} & 0.087 & 0.112 & -0.128 & 0.297 \\ 
 &  \rm{Time} & -0.350 & 0.014 & -0.379 & -0.322 \\ 
 & $\sigma^2$ & 1.017 & 0.062 & 0.900 & 1.144 \\  \hline
 Peri & \rm{Intercept} & -0.074 & 0.116 & -0.310 & 0.132 \\ 
 &  \rm{Time} & 0.605 & 0.029 & 0.549 & 0.661 \\ 
 & $\sigma^2$ & 1.704 & 0.493 & 0.945 & 2.671 \\  \hline
Deep  & \rm{Intercept} & -0.009 & 0.058 & -0.120 & 0.108 \\ 
 &  \rm{Time} & 0.019 & 0.012 & -0.004 & 0.043 \\ 
  & $\sigma^2$ & 0.601 & 0.097 & 0.394 & 0.770 \\  \hline
SBP  & \rm{Intercept} & 0.053 & 0.085 & -0.071 & 0.198 \\ 
 & \rm{Time} & -0.327 & 0.286 & -0.668 & -0.027 \\ 
 & \rm{Time}$^2$ & 0.064 & 0.071 & -0.008 & 0.140 \\ 
& $\sigma^2$ & 0.994 & 0.037 & 0.934 & 1.072 \\  \hline
DBP &  \rm{Intercept} & -0.038 & 0.108 & -0.179 & 0.118 \\ 
&  \rm{Time} & -0.488 & 0.339 & -0.863 & -0.135 \\ 
&   \rm{Time}$^2$ & 0.075 & 0.084 & -0.010 & 0.162 \\ 
&  $\sigma^2$ & 0.967 & 0.040 & 0.907 & 1.062 \\  \hline
BMI &  \rm{Intercept} & 0.014 & 0.225 & -0.273 & 0.299 \\ 
&  \rm{Time} & 0.015 & 0.047 & -0.057 & 0.079 \\ 
&  \rm{Time}$^2$ & 0.003 & 0.017 & -0.015 & 0.022 \\ 
&  $\sigma^2$ & 0.997 & 0.044 & 0.933 & 1.100 \\  \hline
CESDT &  \rm{Intercept} & 0.053 & 0.163 & -0.160 & 0.258 \\ 
 &  \rm{Time} & -0.060 & 0.196 & -0.305 & 0.150 \\ 
 &   \rm{Time}$^2$ & 0.011 & 0.054 & -0.044 & 0.070 \\ 
&  $\sigma^2$ & 1.029 & 0.028 & 0.981 & 1.091 \\ \hline
BENTON &   \rm{Intercept} & 0.110 & 0.063 & 0.010 & 0.235 \\ 
&   \rm{Time} & -0.239 & 0.223 & -0.517 & 0.000 \\ 
&    \rm{Time}$^2$ & 0.049 & 0.053 & -0.005 & 0.107 \\ 
&   $\sigma^2$ & 0.980 & 0.030 & 0.925 & 1.037 \\ \hline
ISA &   \rm{Intercept} & 0.069 & 0.113 & -0.079 & 0.251 \\ 
&   \rm{Time} & -0.143 & 0.229 & -0.436 & 0.109 \\ 
&   \rm{Time}$^2$ & 0.029 & 0.051 & -0.023 & 0.085 \\ 
&  $\sigma^2$ & 0.963 & 0.027 & 0.911 & 1.017 \\ \hline
TOTMED &   \rm{Intercept} & -0.096 & 0.118 & -0.264 & 0.061 \\ 
 &  \rm{Time} & -0.038 & 0.327 & -0.426 & 0.304 \\ 
&    \rm{Time}$^2$ & 0.038 & 0.077 & -0.041 & 0.121 \\ 
&  $\sigma^2$ & 1.053 & 0.046 & 0.977 & 1.140 \\ \hline
TMTA &   \rm{Intercept} & 0.044 & 0.033 & -0.021 & 0.104 \\ 
&  \rm{Time} & -0.193 & 0.245 & -0.500 & 0.067 \\ 
&   \rm{Time}$^2$ & 0.051 & 0.057 & -0.007 & 0.113 \\ 
 &  $\sigma^2$ & 0.957 & 0.049 & 0.879 & 1.058 \\ \hline
TMTB &   \rm{Intercept} & -0.000 & 0.035 & -0.068 & 0.069 \\ 
&   \rm{Time} & -0.250 & 0.205 & -0.492 & -0.031 \\ 
&    \rm{Time}$^2$ & 0.040 & 0.054 & -0.015 & 0.098 \\ 
&  $\sigma^2$ & 0.975 & 0.052 & 0.893 & 1.091 \\ \hline
IADL &  \rm{Intercept} & 0.015 & 0.123 & -0.144 & 0.180 \\ 
&   \rm{Time} & -0.055 & 0.089 & -0.175 & 0.050 \\ 
&   \rm{Time}$^2$ & 0.020 & 0.025 & -0.006 & 0.048 \\ 
&   $\sigma^2$ & 0.969 & 0.032 & 0.915 & 1.038 \\ 
   \hline
\end{tabular}
\end{table}
\FloatBarrier


\begin{table}[ht] 
 \caption{\label{c21} The estimated regression coefficients and variance of errors of the longitudinal sub-model from the joint model in 3C study data using CS. Est: posterior mean, SD: standard deviation, 2.5$\% $ CI: lower bound of credible interval  and  97.5$\% $ CI: upper  bound of credible interval.}
\centering
\scriptsize
\begin{tabular}{ccccccc}
  \hline
 & Est. & SD & $2.5\%$ CI & $97.5\%$ CI & $lBFDR$ & BF  \\ \hline
 &\multicolumn{6}{c}{Dementia} \\ 
 WMV & -0.030 & 0.022 & -0.076 & 0.008 & 0.991 & 0.009  \\ 
 GMV & 0.009 & 0.019 & -0.029 & 0.042 & 1.000 & 0.000  \\ 
 TIV & 0.033 & 0.011 & 0.011 & 0.053 & 1.000 & 0.000 \\ 
 RHIPP & -0.020 & 0.019 & -0.055 & 0.015 & 1.000 & 0.000 \\ 
 LHIPP & -0.042 & 0.021 & -0.085 & -0.001 & 1.000 & 0.000 \\ 
 Peri & 0.021 & 0.015 & -0.005 & 0.053 & 1.000 & 0.000 \\ 
 Deep & 0.020 & 0.026 & -0.037 & 0.065 & 1.000 & 0.000 \\ 
 SBP & -0.006 & 0.040 & -0.078 & 0.062 & 0.988 & 0.012  \\ 
 DBP & -0.041 & 0.032 & -0.102 & 0.020 & 1.000 & 0.000 \\ 
 BMI & -0.003 & 0.014 & -0.032 & 0.027 & 1.000 & 0.000 \\ 
 CESDT & 0.038 & 0.030 & -0.022 & 0.100 & 1.000 & 0.000 \\ 
 BENTON & -0.893 & 0.187 & -1.251 & -0.604 & 0.000 & Inf \\ 
 ISA & -0.034 & 0.031 & -0.094 & 0.027 & 1.000 & 0.000 \\ 
 TOTMED & -0.040 & 0.033 & -0.107 & 0.019 & 1.000 & 0.000 \\ 
 TMTA & 0.004 & 0.042 & -0.077 & 0.074 & 1.000 & 0.000 \\ 
 TMTB & -0.851 & 0.489 & -1.344 & -0.079 & 0.297 & 2.364 \\ 
 IADL & 0.440 & 0.089 & 0.296 & 0.623 & 0.000 & Inf \\ 
 Sex & -0.001 & 0.032 & -0.062 & 0.061 & 1.000 & 0.000 \\ 
 Diabete & 0.009 & 0.054 & -0.054 & 0.071 & 0.979 & 0.021  \\ 
 APOE & 0.688 & 0.183 & 0.311 & 1.020 & 0.000 & Inf \\ 
 Educ & 0.641 & 0.175 & 0.283 & 0.951 & 0.000 & Inf \\ 
 Age & 0.005 & 0.049 & -0.063 & 0.082 & 0.955 & 0.047  \\ 
 &\multicolumn{6}{c}{Death } \\ 
 WMV & 0.016 & 0.023 & -0.030 & 0.061 & 1.000 & 0.000 \\ 
 GMV & -0.010 & 0.019 & -0.046 & 0.026 & 1.000 & 0.000 \\ 
 TIV & 0.029 & 0.011 & 0.008 & 0.051 & 1.000 & 0.000 \\ 
 RHIPP & -0.007 & 0.017 & -0.038 & 0.024 & 1.000 & 0.000 \\ 
 LHIPP & 0.018 & 0.018 & -0.015 & 0.052 & 1.000 & 0.000 \\ 
 Peri & 0.001 & 0.011 & -0.020 & 0.023 & 1.000 & 0.000 \\ 
 Deep & 0.002 & 0.025 & -0.045 & 0.053 & 1.000 & 0.000 \\ 
 SBP & 0.004 & 0.028 & -0.051 & 0.057 & 1.000 & 0.000 \\ 
 DBP & 0.296 & 0.047 & 0.203 & 0.393 & 0.000 & Inf \\ 
 BMI & 0.012 & 0.010 & -0.008 & 0.031 & 1.000 & 0.000 \\ 
 CESDT & -0.003 & 0.025 & -0.050 & 0.050 & 1.000 & 0.000 \\ 
 BENTON & -0.088 & 0.030 & -0.152 & -0.032 & 1.000 & 0.000 \\ 
 ISA & -1.761 & 0.084 & -1.936 & -1.620 & 0.000 & Inf \\ 
 TOTMED & 0.075 & 0.027 & 0.022 & 0.129 & 1.000 & 0.000 \\ 
 TMTA & 0.014 & 0.028 & -0.041 & 0.064 & 1.000 & 0.000 \\ 
 TMTB & 0.014 & 0.027 & -0.041 & 0.068 & 1.000 & 0.000 \\ 
 IADL & 0.011 & 0.025 & -0.040 & 0.058 & 1.000 & 0.000 \\ 
 Sex & -1.816 & 0.196 & -2.088 & -1.813 & 0.000 & Inf \\ 
 Diabete & 0.125 & 0.221 & -0.050 & 0.662 & 0.763 & 0.311  \\ 
 APOE & 0.002 & 0.039 & -0.063 & 0.067 & 0.979 & 0.021  \\ 
 Educ & 0.001 & 0.032 & -0.062 & 0.067 & 1.000 & 0.000 \\ 
 Age & 1.227 & 0.170 & 0.864 & 1.543 & 0.000 & Inf \\ 
\hline
\end{tabular}
\end{table}

\end{document}